\documentclass[%
prx, 
 amsmath,amssymb,
reprint,%
]{revtex4-2}

\usepackage{float}
\usepackage{graphicx}
\usepackage{dcolumn}
\usepackage{bm}
\usepackage[mathlines]{lineno}

\usepackage[utf8]{inputenc}
\usepackage[T1]{fontenc}
\usepackage{mathptmx}
\usepackage{etoolbox}

\usepackage{graphicx}
\usepackage{dcolumn}
\usepackage{bm}
\usepackage{comment}
\usepackage{hyperref}
\usepackage{import}
\usepackage[normalem]{ulem}
\usepackage{ mathrsfs }
\usepackage{ bbold }
\usepackage{ textcomp }
\usepackage{ stmaryrd }
\usepackage{float}
\usepackage{physics}
\usepackage{yhmath}
\usepackage{bm}
\usepackage{xfrac}
\usepackage{tabularx}
\usepackage{multirow}
\usepackage[dvipsnames, table]{xcolor}
\usepackage{algorithm}
\usepackage[noend]{algpseudocode}
\usepackage{lineno}

\definecolor{redd}{rgb}{0, 0, 0}

\definecolor{red}{rgb}{0, 0, 0}

\newcommand{\notes}[1]{\textcolor{black}{#1}}
\newcommand{\nnotes}[1]{\textcolor{black}{#1}}
\newcommand{\chg}[1]{\textcolor{redd}{#1}}
\newcommand{\new}[1]{\textcolor{black}{#1}}
\newcommand{\st}[1]{\iffalse{#1}\fi}
\newcommand{\stt}[1]{\iffalse{#1}\fi}
\newcommand\so[1]{\iffalse{#1}\fi}

\newcommand{\citex}[1]{\cite{#1} (\citeyear{#1}) }

\usepackage[english]{babel}

\makeatletter
\def\bbl@set@language#1{%
  \edef\languagename{%
    \ifnum\escapechar=\expandafter`\string#1\@empty
    \else\string#1\@empty\fi}%
  \@ifundefined{babel@language@alias@\languagename}{}{%
    \edef\languagename{\@nameuse{babel@language@alias@\languagename}}%
  }%
  \select@language{\languagename}%
  \expandafter\ifx\csname date\languagename\endcsname\relax\else
    \if@filesw
      \protected@write\@auxout{}{\string\select@language{\languagename}}%
      \bbl@for\bbl@tempa\BabelContentsFiles{%
        \addtocontents{\bbl@tempa}{\xstring\select@language{\languagename}}}%
      \bbl@usehooks{write}{}%
    \fi
  \fi}
\newcommand{\DeclareLanguageAlias}[2]{%
  \global\@namedef{babel@language@alias@#1}{#2}%
}
\makeatother

\DeclareLanguageAlias{en}{english}

\makeatletter
\def\@email#1#2{%
 \endgroup
 \patchcmd{\titleblock@produce}
  {\frontmatter@RRAPformat}
  {\frontmatter@RRAPformat{\produce@RRAP{*#1\href{mailto:#2}{#2}}}\frontmatter@RRAPformat}
  {}{}
}%
\makeatother
\begin{document}

\preprint{arXiv:2302.04543}

\title{Noisy Qudit vs Multiple Qubits  : Conditions on Gate Efficiency \chg{for Enhancing Fidelity}}
\author{Denis Jankovi\'c}
\email{denis.jankovic@ipcms.unistra.fr}
\affiliation{Universit\'e de Strasbourg, CNRS, Institut de Physique et Chimie des Mat\'eriaux de Strasbourg (IPCMS), UMR 7504, 23, Rue du Loess, 67000 Strasbourg, France}
\affiliation{Institute of Nanotechnology (INT), Karlsruhe Institute of Technology, P.O. Box 3640, 76021 Karlsruhe, Germany}

\author{Jean-Gabriel Hartmann}
\email{jean-gabriel.hartmann@ipcms.unistra.fr}
\affiliation{Universit\'e de Strasbourg, CNRS, Institut de Physique et Chimie des Mat\'eriaux de Strasbourg (IPCMS), UMR 7504, 23, Rue du Loess, 67000 Strasbourg, France}

\author{Mario Ruben}
\email{mario.ruben@kit.edu}
\affiliation{Institute of Nanotechnology (INT), Karlsruhe Institute of Technology, P.O. Box 3640, 76021 Karlsruhe, Germany}
\affiliation{Institute for Quantum Materials and Technologies (IQMT), Karlsruhe Institute of Technology, P.O. Box 3640, 76021 Karlsruhe, Germany}
\affiliation{Centre Europ\'een de Sciences Quantiques (CESQ), Institut de Science et d'Ing\'enierie Supramol\'eculaires (ISIS), Universit\'e de Strasbourg, 8, All\'ee Gaspard Monge, 67000 Strasbourg, France}

\author{Paul-Antoine Hervieux}
\email{hervieux@unistra.fr}
\affiliation{Universit\'e de Strasbourg, CNRS, Institut de Physique et Chimie des Mat\'eriaux de Strasbourg (IPCMS), UMR 7504, 23, Rue du Loess, 67000 Strasbourg, France}

\date{\today}

\begin{abstract}
\notes{As qubit-based platforms face near-term technical challenges in terms of scalability}\st{Today}, \textit{qudits}, $d$-level quantum bases of information, are being implemented in multiple platforms \notes{as an alternative} for Quantum Information Processing (QIP). {It is, therefore, crucial to study their efficiencies for QIP compared to more traditional qubit platforms}\notes{, specifically since each additional quantum level represents an additional source of environmental coupling}. {We} present a comparative study of the infidelity scalings of a qudit and $n$-qubit systems, both with identical Hilbert space dimensions and noisy environments. The first-order response of the \textit{Average Gate Infidelity} (AGI) to the noise in the Lindblad formalism, which was found to be gate-independent, was {calculated analytically} in the two systems being compared. \stt{This yielded a critical curve $O(d^2/\log_2(d))$ of the ratio of their respective gate times in units of decoherence time}This yielded a critical curve \new{$(d^2-1)/3\log_2(d)$} of the ratio of their respective \st{gate times in units of decoherence time} \new{figure of merits, defined as the gate time in units of decoherence time.} \new{This quantity indicates how time-efficient operations on these systems are \nnotes{relative to decoherence timescales}, and the critical curve is especially useful for precisely benchmarking qudit platforms with smaller values of $d$}. The curve delineates regions where each system has a higher rate of increase of the AGI than the other. This condition on gate efficiency was applied to different existing platforms. \notes{Specific qudit platforms were found to} possess gate efficiencies competitive with state-of-the-art qubit platforms. Numerical simulations complemented this work and allowed for discussion of the applicability and limits of the linear response formalism.


\end{abstract}

\maketitle



\section*{\label{sec:intro} Introduction}

The paradigmatic bases of information in Quantum Information Processing (QIP) are  qubits: two-level individually addressable quantum systems. {However, several QIP platforms have recently been proposed that instead make use of $d$-level systems, referred to as qudits} \chg{\cite{moreno-pineda_molecular_2018, PhysRevB.95.064423, chi_programmable_2022, ringbauer_universal_2022,PhysRevApplied.19.064060,cao2023emulating}}. In its infancy, classical computing did experiment with ternary, quaternary, or higher-dimensional, bases of information, before eventually settling on the simplest (bits), when near-zero error rates and easy scalability were attained \cite{brusentsov_ternary_2011}. Analogously, it could be argued that quantum computing is likely to follow a similar trend in the long-term; as fault-tolerant platforms emerge and technologies mature, the industry could indeed fully settle on multiqubit systems. However, QIP research is currently not in the noise-free regime but near it, and in order to reach significant quantum supremacy \cite{boixo_characterizing_2018}, increasing the total Hilbert {space dimension} of the physical platform is a primordial requirement. As such, there is a current race to increase the number $n$ of coupled qubits ($d=2$) with superconducting platforms leading the way with $n=51$ \cite{arute_quantum_2019} or $n=433$ \cite{noauthor_ibm_2022}. While in general the Hilbert {space dimension} increases exponentially in the number of sites, the relatively slow $2^n$ scaling of qubits, compared to $d^n$ for qudits, is proving challenging, necessitating ever-more robust systems and complex control mechanisms. 

Given these current technical challenges, most qudit-based platforms that have been physically implemented argue for near-term advantages over equivalent multiqubit implementations. Thus, the principal motivations of qudit platforms over qubits include: (i) the underlying physical systems having lower decoherence rates \cite{wang_qudits_2020}, (ii) using the redundancy in additional levels for quantum error correction \cite{chiesa_embedded_2021, petiziol_counteracting_2021}, (iii) the higher density of information per physical system (site) \cite{wernsdorfer_synthetic_2019}\chg{, (iv) the reduced number of nonlocal, hence more decoherence-sensitive, operations \cite{Luis2020}} or (\so{i}v) more robust flying quantum memories \cite{zheng_entanglement_2022, bouchard_high-dimensional_2017}. Furthermore, qudits present fundamental theoretical advantages, enabling novel QIP capabilities offered by $\bigotimes SU(d)$ vs. $\bigotimes SU(2)$ of qubits \cite{campbell_magic-state_2012} \chg{such as simplifying some quantum algorithms \cite{Lanyon2008}}, and therefore a fault-tolerant qudit quantum computer indeed remains conceivable. Hence, qudits provide an alternative scaling solution by linearly increasing $d$ \nnotes{- instead of scaling up $n$, the number of sites}, as well as increasing efficiency through single qudit gates operating on larger computational subspaces \cite{arute_quantum_2019,ringbauer_universal_2022,godfrin_operating_2017,thiele_electrically_2014}. However, one of the disadvantages raised for qudits is the larger number of error channels compared to multiple qubits \cite{otten_impacts_2021}. In this context, a study of the near-term viability of qudits is needed to investigate the interplay between computational efficiency and noise error rates in higher dimensions.

In this work, \notes{we consider} \nnotes{one} single qudit \nnotes{versus} multiqubit systems, in the context of near noise-free implementations. \nnotes{We undertake an inquiry to determine under what conditions on the applied gates a single qudit system does not lose more computational information than an equivalent multiqubit system, even when the qudit system initially presents more potential error channels. For this purpose, a standard measure to quantify the loss of computational information, that we} study\nnotes{, is} the Average Gate Infidelities (AGI)\notes{, as defined by \citet{nielsen_simple_2002} where the average is over the Haar measure.} \nnotes{The choice of the AGI ensures the calculated fidelity is not dependent on the input state and therefore remains relevant even if the gate is applied in later stages of a quantum algorithm.} \st{Comparing systems of equivalent Hilbert space dimension undergoing arbitrary unitary transformations, we investigate the respective growth rates of the AGIs with respect error rates $\gamma$ and dimensionless gate time $\gamma t$.}\notes{We conduct an in-depth analysis in which we compare the computational fidelity of a single qudit and $n$-qubit systems, both with identical dimensions of Hilbert space, undergoing arbitrary unitary transformations, and evolving under the influence of comparable noisy conditions. Our benchmark for successful analysis is defined by a lower first-order response of the AGI to the environmental noise, providing a measure of computational fidelity independent of initial states. We investigate\nnotes{, for increasing values of $d$,} the respective growth rates of the AGIs with respect to error rates $\gamma$ and dimensionless gate time $\gamma t$.} \nnotes{The latter quantifies the gate efficiency by indicating how time-efficient operations on these systems are relative to decoherence timescales, therefore this paper presents a study of the first-order connection between the AGI and this time-efficiency $\gamma t$.} 

\nnotes{In other words, this study aims to investigate how the AGI scales proportionally with both the error rate and the speed at which gate operations are performed, as well as the dimension $d$ of the qudit. Additionally, this study aims to provide a benchmarking tool \so{for} \chg{to decide} if a qudit platform, for a given ($\gamma$,$t$,$d$) specification, can compensate for its greater number of \so{potential} error channels by leveraging advantageous decoherence times and gate speeds. Both of those quantities depend intrinsically on the physical platform implementing the single qudit or multiqubit system, in particular their coupling to the environment, the mathematical form of the control pulses' Hamiltonian, and the addressing speed. In particular, given a single qudit platform and a multiqubit platform with equivalent Hilbert space dimensionality, and specifying a fixed pair of parameters ($\gamma$,$t$), one could conduct a comparative analysis to determine if the qudit platform exhibits sufficiently low decoherence and sufficiently rapid gate time to achieve computational fidelities that are competitive with the multiqubit platform. Or, similarly, since increasing $d$ on a single site in a given qudit platform is a prevailing goal for some platforms \cite{wernsdorfer_synthetic_2019}, assuming $\gamma t$ remains of the same order of magnitude, this study would also allow setting theoretical upper limits on the value of $d$ in order to remain advantageous.}

In \so{sec.\ref{sec:2}}\chg{the first part,} a gate-independent formula is presented for the first-order response in $\gamma t$ of the AGI to Markovian noise in the Lindblad formalism. The first-order formalism corresponds to the quasi-errorless regime of near-term QIP systems. Expressions for the linear dependency of the AGI \so{in} \chg{on} $\gamma t$ for a single qudit, multiqubits and {also} multiqudits are derived for an arbitrary collapse operator. A comparison is then made between the rate of increase of the AGI of a single qudit vs. equivalent multiple qubits.

\so{In sec.\ref{sec:3}} \chg{This is then followed by} numerical simulations, {performed} with the \texttt{Python} package \texttt{QuTiP} \cite{johansson_qutip_2012}, \chg{that} complement {and illustrate} the analytical results. Discussions of the applicability and limits of the linear response formalism for AGI {are given and the following aspects are studied}: (i) the applicable range of $\gamma t$ and its dependency on the dimension of the qudit; (ii) the extent of the gate-independence of the result; (iii) the applicability to noise models other than pure dephasing; and finally (iv) the conditions on gate times for which either qudits or multiple qubits are advantageous. {This latter aspect is then examined in more detail with respect to existing platforms \nnotes{by taking into account their respective decoherence rate and gate operation time}.}



\section*{Results and Discussion}

\subsection*{\label{ssec:2:fluc-dissip} Fluctuation-dissipation relation \chg{for a perturbed pure state}}

Consider a qudit, a $d$-level quantum system whose dynamics are governed by the Lindblad master equation \cite{manzano_short_2020}:
\begin{equation}
\label{eq:master}
    \frac{\text{d}\\rho}{\text{d}t} = -i\left[H,\rho\right] + \sum_{k=1}^K \gamma_k \left( L_k \rho L_k^\dag  - \frac{1}{2} \left\{ L_k^\dag L_k , \rho \right\} \right),
\end{equation}
{where} $\rho(t)$ {is} the density matrix of the system at time $t$, $H$ the Hamiltonian of the system, $L_k$ the so-called \textit{collapse operators} characterizing the Markovian noise, {and $\gamma_k$ the decay parameters} for each of the $K$ noise processes. \chg{$H=H_0+H_c(t)$ where $H_0$ models the free evolution of the physical system and encompassing its internal interactions, and $H_c(t)$ is a time-dependent pulse Hamiltonian allowing the controlled evolution. Moreover, the interactions of $H$ with the collapse operators determine relevant timescales such as the gate-time $t$ and the decoherence time $T_2$ that are thus inherent to the physical realization under consideration.}

The aim is to study the effect of a single collapse operator $\sqrt{\gamma_1} L_1 =\sqrt\gamma L$ on short timescales and under small-amplitude noise, i.e., $\gamma t \ll 1$. Under these assumptions, one can consider an \textit{ansatz} of the form: 
\begin{equation}
\label{eq:rhotpert}
    \rho(t) = \rho^* - \gamma t M +\chg{\mathcal{O}( (\gamma t)^2 )},
\end{equation}
with $\rho^*$ the noiseless target state, which is the solution of $\dot{\rho} = -i\left[H,\rho\right]$ after time $t$, and $M$ the perturbation matrix resulting from the presence of a small-amplitude noise. \chg{Terms in $\chg{\mathcal{O}( (\gamma t)^2 )}$ include terms whose prefactor is of the form $\left(\gamma^l t^k\right)_{l+k\geq3}$, a more in-depth discussion is available in Appendix \ref{apdx:exp}.}

{One can easily see that the use of (\ref{eq:rhotpert}) in (\ref{eq:master}) leads to}
\begin{align}
\label{eq:rhopertresult}
M = \frac{1}{2} \left\{ L^\dag L, \rho^* \right\} - L \rho^* L^\dag.
\end{align}

Consider now a quantum operation bringing the initial state $\rho_0$ to a final state $\rho(t)$ at time $t$. One define the \textit{fidelity} $\mathcal{F}$ of this final state relative to some target state $\rho^*$ \cite{jozsa_fidelity_1994} as
\begin{equation}
    \label{eq:deffid}
    \mathcal{F}(\rho(t),\rho^*) \equiv \left[\Tr\left(\sqrt{\sqrt{\rho(t)}\rho^*\sqrt{\rho(t)}}\right)\right]^2. 
\end{equation}
Subsequently, the \textit{infidelity}, is then defined as
\begin{equation}
    \label{eq:deferr}
    \mathscr{E} \equiv 1-\mathcal{F}.
\end{equation}
Since $\rho^*$ is a pure state ($\rho^* = \ket{\varphi^*}\bra{\varphi^*}$) Eq.(\ref{eq:deffid}) simplifies to \cite{jozsa_fidelity_1994}
\begin{equation}
    \label{eq:deffidsimpl}
    \mathcal{F}(\rho(t),\rho^*) = \Tr\left(\rho(t)\rho^*\right). 
\end{equation}
{Finally, substituting} Eq.(\ref{eq:rhotpert}) into Eq.(\ref{eq:deffidsimpl}) leads to (see \ref{apdx:flucdissip})
\begin{equation}
\label{eq:flucdissip}
    \mathscr{E}(\rho^*) = \gamma t \Delta_* L+ \chg{\mathcal{O}( (\gamma t)^2 )}. 
\end{equation}
{where $\Delta_* L= \langle L^\dag L\rangle_{*} - \langle L^\dag \rangle_*\ev{L}_* $ with $\langle L^\dag L\rangle_{*} \equiv \Tr\left(\rho^*L^\dag L\right)$.}

\subsection*{\label{ssec:2:avgafid} Average Gate Fidelity \chg{of a single qudit}}

Only $\mathscr{E}(\rho^*)$ for a specific $\rho^*$ was obtained \so{beforehand} \chg{in the previous subsection}. However, is there a state-independent approach to obtaining the infidelity of a quantum gate under small-amplitude noise? One defines the quantum gate $U$ applied {during a time duration} $t$, whose resulting operation brings all initial states $\rho_0$ to all corresponding $\rho^* = U\rho_0U^\dag$. There is then a definition of the \textit{average gate fidelity} of a quantum channel $\mathcal{E}$\chg{,} attempting to \so{implement} \chg{carry} the unitary operation $U$ \so{with} \chg{despite} a noisy environment\chg{,} {which reads as follows} \cite{nielsen_simple_2002}
\begin{equation}
    \label{eq:avgatefid}
    \begin{aligned}
    \bar{\mathcal F}(\mathcal{E}, U) &= \int d \rho_0 \, \mathcal{F}({\rho(t)},\rho^*)\\&= \int d \rho_0 \, \left\langle U^{\dagger} \mathcal{E}[\rho_0] U\right\rangle_0 = \int d \rho_0 \,\left\langle\left(\mathcal{U^\dag \circ E}\right)[\rho_0]\right\rangle_0,
    \end{aligned}
\end{equation}
where the normalized integral is over the Fubini-Study measure  on pure states (sometimes called the Haar measure) \cite{qi_comparing_2019}, $\mathcal{U}^{\dagger}[\rho] \equiv U^{\dagger} \rho U$ and $\mathcal{E}[\rho_0] = \rho(t)$.

Introducting $\tilde E_k = E_k U$ the \textit{Kraus operators} such that
\begin{equation}
    \rho(t) = \left(\mathcal{U^\dag \circ E}\right)[\rho_0] = \sum_k \tilde E_k \rho_0 \tilde E_k^\dag = \sum_k E_k \rho^* E_k^\dag,
\end{equation}
{the Average Gate Fidelity $\bar{\mathcal F}$ given in (\ref{eq:avgatefid}) can be rewritten as \cite{johnston_quantum_2011, magesan_gaining_2008}}
\begin{equation}
\label{eq:krausfid}
\bar{\mathcal F}(\mathcal{E}, U) = \frac{d + \sum_k\left|\Tr\left(\tilde E_k U^\dag\right)\right|^2}{d(d+1)} = \frac{d + \sum_k\left|\Tr\left(E_k\right)\right|^2}{d(d+1)}\;.
\end{equation}

Using Eq.(\ref{eq:rhopertresult}), one seeks {the sets of Kraus operators $\left\{\tilde E_k\right\}$ or $\left\{ E_k \right\}$} such that, to $\chg{\mathcal{O}( (\gamma t)^2 )}$, $\forall \rho_0, \rho^*,$
    \begin{equation}
\label{eq:krauslindblad}
        \begin{aligned}\sum_k \tilde E_k \rho_0 \tilde E_k^\dag &= U\rho_0U^\dag - \gamma t \frac{1}{2} \left\{ L^\dag L, U\rho_0U^\dag \right\} + \gamma t L U\rho_0U^\dag L^\dag\\ 
        \quad \sum_k E_k \rho^* E_k^\dag &= \rho^* - \gamma t \frac{1}{2} \left\{ L^\dag L, \rho^* \right\} + \gamma t L \rho^* L^\dag. 
        \end{aligned}
    \end{equation}

One can see that the following two sets would work up to the first order in $\gamma t$

\begin{equation}
\label{eq:krausop}
\begin{aligned}
    \tilde E_0 &= \left(\mathbb{1}_d - \frac{\gamma t}{2} L^\dag L\right)U, &\tilde{E_1} =\sqrt{\gamma t} LU, \\
    E_0 &= \mathbb{1}_d - \frac{\gamma t}{2} L^\dag L, &E_1 =\sqrt{\gamma t} L.\phantom{U}
\end{aligned}
\end{equation}

In order to use Eq.(\ref{eq:krausop}) in Eq.(\ref{eq:avgatefid}), it is necessary to calculate the traces of the operators. {Let us consider} a pure dephasing channel \notes{of a qudit coupled to a thermal environment through the operator $J_z$ (In general the coupling of a qudit, or qubit, to a thermal environment can be represented by a linear combination, or mixture, of collapse operators, though a pure dephasing channel will typically be present and can be represented by the operator $J_z$. As a toy model, let us consider a coupling term dominated by a pure dephasing channel.)} i.e. $\mathcal{E}_z$ with $L=J_z$ \cite{zhong_fisher_2013} . One obtains a {\it gate}- (and Hamiltonian-) independent result {for the Average Gate Fidelity which reads as (see \ref{apdx:puredephkraus})}
\begin{equation}
\label{eq:Fid112}
    \overline{\mathcal{F}}(\mathcal{E}_z) = 1-\frac{\gamma t}{12} d(d-1) + \chg{\mathcal{O}( (\gamma t)^2 )}. 
\end{equation}

In other words the \textit{Average Gate Infidelity} \chg{(AGI)} is given by
\begin{equation}
\label{eq:avEqudits}
    \overline{\mathscr{E}}(\mathcal{E}_z) = \frac{\gamma t}{12} d(d-1) + \chg{\mathcal{O}( (\gamma t)^2 )}, 
\end{equation}
{or more generally} for an arbitrary quantum channel $\mathcal{X}$ with collapse operator $L$
\begin{equation}
\label{eq:avEquditsL}
    \overline{\mathscr{E}}(\mathcal{X}) =  \frac{\gamma t}{d+1} \left( \Tr(L^\dag L) - \frac{1}{d}\left|\Tr(L)\right|^2 \right) + \chg{\mathcal{O}( (\gamma t)^2 )}. 
\end{equation}

{Note that it is always possible to find a traceless collapse operator $L$ emulating $\mathcal{X}$ \cite{manzano_short_2020}, so the previous expression can be, in this case, simplified as follows}
\begin{equation}
\label{eq:avEquditsLsimp}
    \overline{\mathscr{E}}(\mathcal{X}) =  \frac{\gamma t}{d+1} \Tr(L^\dag L) + \chg{\chg{\mathcal{O}( (\gamma t)^2 )}}. 
\end{equation}

It follows from (\ref{eq:avEquditsLsimp}) that\chg{, if $L$ were independent of $d$,} increasing the dimension $d$ {of the Hilbert space} would also increase the {\it robustness} of qudit gates to a dimension-independent quantum channel \so{({if $L$ is} independent of $d$)}. 

\subsection*{\label{ssec:2:bvsd}\chg{AGI of} qudits vs. qubits}

{Now let us apply the same technique as described above to another system:} an ensemble of $n$ identical dephasing qubits (Hilbert space of dimension $d=2^n$). \notes{In order to compare it with the qudit analysis in the previous subsection, each individual qubit decoheres with the same rate (has the same type, and strength, of environmental coupling) through its spin operator $S_z$ in the same way as the individual qudit (with $d=2$). Considering any additional coupling mechanism to the environment arising from inter-qubit interactions would only further disadvantage the multi-qubit implementation. Our considerations then provide a best-case scenario for comparable qubits.} \st{yielding}\notes{This yields} the master equation
\begin{equation}
    \frac{\text{d}\rho}{\text{d}t} = -i\left[H,\rho\right] + \sum_{k=1}^n L_k \rho L_k^\dag  - \frac{1}{2} \sum_{k=1}^n \left\{ L_k^\dag L_k , \rho \right\} \;,
\end{equation}
with
\begin{equation} L_k = \underbrace{\mathbb{1}_2^{(1)} \otimes \dots \otimes \mathbb{1}_2^{(k-1)}}_{k-1}\otimes S_z^{(k)} \otimes \underbrace{\mathbb{1}^{(k+1)}_2 \otimes \dots \otimes \mathbb{1}^{(n)}_2}_{n-k} \label{eq:Lkqubit}
\end{equation}
for $k\in\llbracket1,n\rrbracket$.

Using the same reasoning as for dephasing qudits one obtains $n+1$ Kraus operators to first order in $\gamma t$
\begin{align}
    \label{eq:krausopqubits}
    E_0 = \mathbb{1}_{2^n} - \sum_{k=1}^n\frac{\gamma t}{2} L_k^\dag L_k, \;\;\;\;\; E_k =\sqrt{\gamma t} L_k.
\end{align}

In this case (see \ref{apdx:agiqubits}),
\begin{equation}
\label{eq:avEqubits}
    \overline{\mathscr{E}}(\mathcal{E}_z) = \frac{\gamma t}{4} \frac{n2^{n}}{2^n+1} + \chg{\mathcal{O}( (\gamma t)^2 )} = \frac{\gamma t}{4} \frac{\log_2(d)d}{d+1} + \chg{\mathcal{O}( (\gamma t)^2 )} \;.
\end{equation}
{Let us stress} that (\ref{eq:avEqubits}) yields the same result as \citet{abad_universal_2022} in the case of identically dephasing qubits with no energy relaxation.

{The analytical expressions (\ref{eq:avEquditsLsimp}) and (\ref{eq:avEqubits}) are one of the main results of this work.}

Following those last two results, two expressions for the AGI have been found: for a single qudit, one finds an infidelity that scales as $d^2$: $\overline{\mathscr{E}_d}(\mathcal{E}_z) = c_d \gamma t$ in (\ref{eq:avEqudits}) and for an ensemble of $n$ qubits one finds an infidelity that scales as {$\log_2(d)$}: $\overline{\mathscr{E}_{b, n}}(\mathcal{E}_z) = c_{b,n} \gamma t$ in (\ref{eq:avEqubits}). Moreover, in the case of pure dephasing, one can define the $T_{2,d}$ dephasing time between two energy-adjacent levels for a qudit. It then shares the same expression (in terms of $\gamma$) as the typical $T_{2,b}$ dephasing time of a single qubit, namely $\dfrac{1}{T_2}=\dfrac{\gamma}{2}$.

The ratio between two average {gate} infidelities, of duration $t_d$ and $t_{b,n}$ for qudit and $n$ qubits, respectively, becomes
\begin{equation}
    \label{eq:ratioE}
    \frac{\overline{\mathscr{E}_d}(\mathcal{E}_z)}{\overline{\mathscr{E}_{b,n}}(\mathcal{E}_z)} = \frac{c_{d} t_{d}/T_{2,d}}{c_{b,n} t_{b,n}/T_{2,b}}.
\end{equation}

Therefore, in order for a {{\it single}} qudit ($d=2^n$) to outperform {{\it an ensemble}} of $n$ qubits in noise-robustness, i.e., to have a smaller AGI, the following {inequality} must hold true
\begin{equation}
    \label{eq:conditionT2}
    \frac{ t_{b,n}/T_{2,b} }{ t_{d}/T_{2,d} } > \frac{c_{d}}{c_{b,n}} = \frac{d^2-1}{3 \log_2(d)} = \frac{4^{n}-1}{3 n}.
\end{equation}

This expression quantifies the requirements on the \textit{figure of merit} which is the gate time in units of decoherence time $\tau_{d} = t_d/T_{2,d}$ relative to $\tau_{b,n} = t_{b,n}/T_{2,b}$ in order for the qudit to yield higher-fidelity gates. Moreover, it confirms that the infidelity of an ensemble of $n$ identical qubits and the infidelity of a single qudit will generally not have the same linear behaviour in $\gamma t$ even if they have the same $T_2$, thus simply having $  \tau_d< \tau_{b,n}$ is not sufficient to guarantee a more noise-resilient qudit. \new{Moreover, Eq.\ref{eq:conditionT2} provides a more precise condition on the ratio of figure of merits than a simple qualitative result such as $\frac{d^2}{\log_2(d)}$, while maintaining the expected $O(d^2/\log_2(d))$ behaviour as $d\rightarrow\infty$. In particular, see \ref{apdx:puredephkraus} for the full analytical calculations, including the derivation of the non-trivial factor $\frac{1}{3}$}.

On a side note, the previous calculations could also be applied to an ensemble of $N$ qudits under identical pure dephasing, in which case {we have}
\begin{equation} \label{eq:d^N}
    \overline{\mathscr{E}_{d,N}}(\mathcal{E}_z) = \frac{\gamma t}{12}\frac{Nd^{N}}{d^N+1}(d^2-1)+ \chg{\mathcal{O}( (\gamma t)^2 )},
\end{equation}
and for $2^n = d^N$ {one obtains} 
\begin{equation}
    \label{eq:conditionT2d^N}
    \frac{ t_{b,n}/T_{2,b} }{ t_{d,N}/T_{2,d} } > \frac{c_{d,N}}{c_{b,n}} = \frac{d^2-1}{3 \log_2(d)}.
\end{equation}

{Let us also note that} for $L$ arbitrary (\ref{eq:d^N}) yields
\begin{equation}
 \label{eq:d^Ngene}
    \overline{\mathscr{E}_{d,N}}(\mathcal{X}) = \gamma t N \frac{d^{N-1}}{d^N+1} \left( \Tr(L^\dag L) - \frac{1}{d}\left|\Tr(L)\right|^2 \right) + \chg{\mathcal{O}( (\gamma t)^2 )}.
\end{equation}

\begin{figure}[htbp!]
    \centering
    \includegraphics[width=\columnwidth]{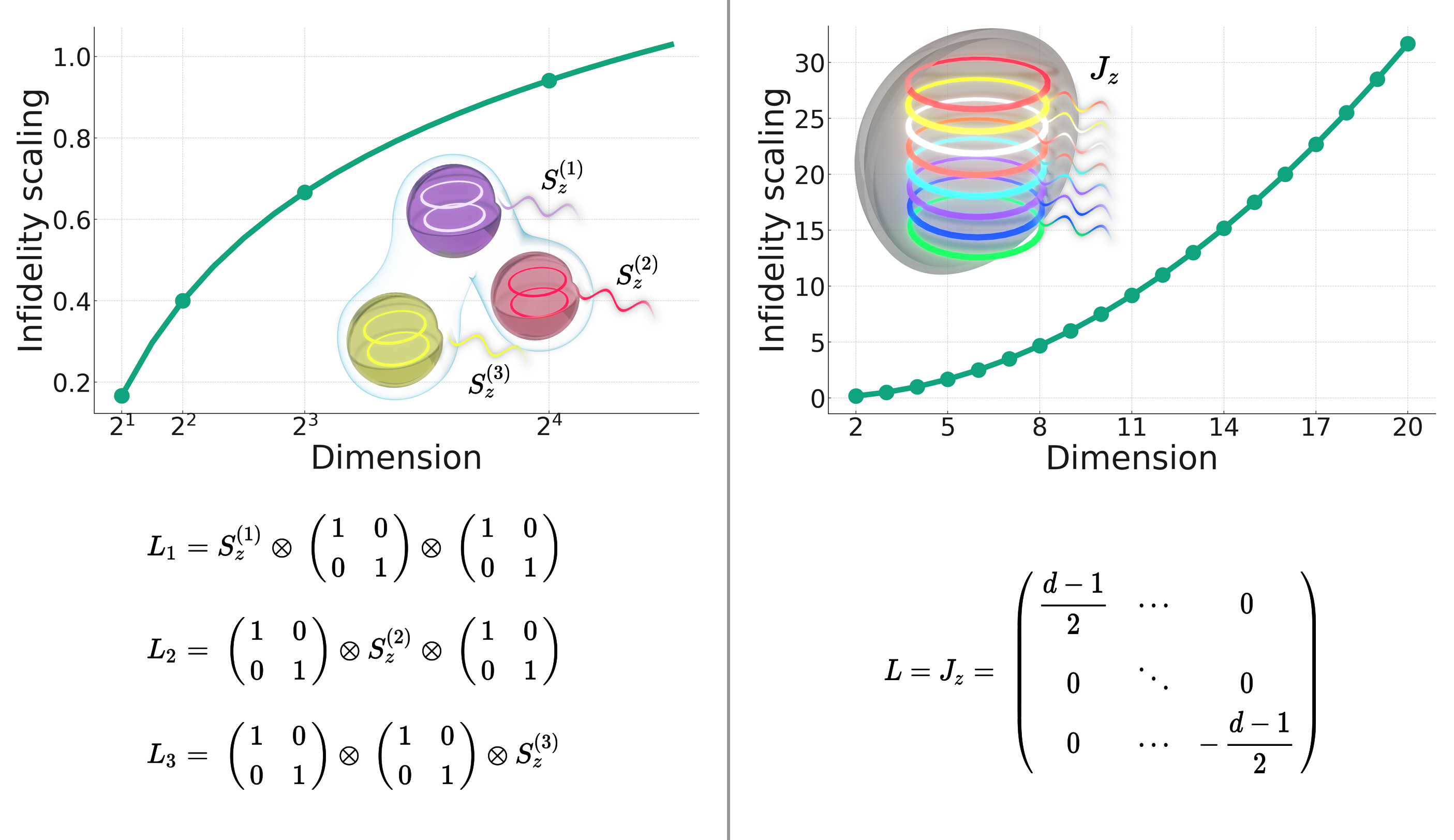}
    \caption{\chg{Summary diagram illustrating the selected collapse operators and the associated analytically derived expected infidelity scalings as functions of the Hilbert Space dimension, as derived from (\ref{eq:avEqubits}) and (\ref{eq:avEqudits}). This is depicted for two distinct systems: multiple qubits (left) and a single qudit (right). The term 'infidelity scaling' here refers to the slopes of the first-order-in-$\gamma t$ AGIs, denoted as $c$ in (\ref{eq:ratioE}).}}
\label{fig:resume}
\end{figure}

\chg{This equation encompasses both scenarios under investigation in this paper until this point. We recall that we construct two systems of equivalent Hilbert Space dimension: a single qudit of dimension $d$ and a system of $N$ qubits. 
(\ref{eq:d^Ngene}) reflects two different scalings of the AGI in $N$ and $d$ respectively. In $N$, it is linear, and, in $d$, it scales as $\left( \Tr(L^\dag L) - \frac{1}{d}\left|\Tr(L)\right|^2 \right)$ since the dimension affects the definition of $L$. In \ref{apdx:puredephkraus} we computed that spin-based $L$ lead to a quadratic scaling in $d$. Therefore, in \hyperref[ssec:2:avgafid]{the qudit subsection}, for a single qudit, we study this quantity for fixed $N=1$ and varying $d$, while in \hyperref[ssec:2:bvsd]{the multiqubits subsection}, it is for fixed $d=2$, but varying $N$. The subtlety in the latter case being that, by construction $N:=\log_2(d)$, with $d$ being that of the single qudit, hence the different scalings in $d$ in (\ref{eq:avEqudits}) and (\ref{eq:avEqubits}). See Fig.\ref{fig:resume} for a visual summary of the results.}

\notes{And if each qudit $k$ has a different set of noise parameters ($\gamma_k$, $L_k$), an even more general formula arises :
\begin{equation}
    \overline{\mathscr{E}_{d,N}}(\mathcal{X}) = \frac{d^{N-1}}{d^N+1} \sum_{k=1}^N \gamma_k t \left( \Tr(L_k^\dag L_k) - \frac{1}{d}\left|\Tr(L_k)\right|^2 \right) + \chg{\mathcal{O}( (\gamma_k t)^2 )}.
\end{equation}}

\notes{This formula through its general form, can be applied to any qudits whose physical implementation implies different collapse operators from the ones considered in this paper.}
\subsection*{\label{ssec:4:processerror} Process fidelity \& averaged fluctuation-dissipation relation}

One may link the fluctuation-dissipation relation obtained in (\ref{eq:flucdissip}) with the results regarding average gate infidelities from Eq. (\ref{eq:avEquditsL}),
\begin{equation}
    \overline{\mathscr{E}}(\mathcal{X}) = {\gamma t}\int d\rho^* \mathscr{E}(\rho^*) = {\gamma t}\int d\rho^* \Delta_* L + \chg{\mathcal{O}( (\gamma t)^2 )}.
\end{equation}

This integral over the Fubini-Study measure can formally be computed using Weingarten calculus methods \cite{collins_weingarten_2021} (see \ref{apdx:haarweingarten}) {which can be expressed as} 
\begin{equation} \label{eq:haarmeanfull}
    \int d\rho \Delta L =  \frac{1}{d+1} \Tr(L^\dag L) - \frac{1}{d(d+1)}\left|\Tr(L)\right|^2,
\end{equation}
{leading to} (\ref{eq:avEquditsL}). 

In contrast to this formal approach, a more physically-informed approach to obtain the same result was proposed in the previous subsections.

Furthermore, it is possible to express all the computed average gate infidelities as process/entanglement infidelities $\mathscr{E}^\text{(p)}$ making use of the relation $D\mathscr{E}^\text{(p)} = (D+1)  \overline{\mathscr{E}}$,  with $D=d$, $2^n$ or $d^{{N}}$ the dimension of the Hilbert space \cite{horodecki_general_1999}. This yields the expression
\begin{equation}
    \mathscr{E}^\text{(p)}_{d,n}(\mathcal{X}) = \gamma t \frac{n}{d} \left( \Tr(L^\dag L) - \frac{1}{d}\left|\Tr(L)\right|^2 \right) + \chg{\mathcal{O}( (\gamma t)^2 )},
\end{equation}
which is linear in the number of subsystems $n$. Likewise we have

\begin{equation}
    \mathscr{E}^\text{(p)}_{d}(\mathcal{E}_z) = \frac{\gamma t}{12} (d^2-1) + \chg{\mathcal{O}( (\gamma t)^2 )}, 
\label{eq:process_Jz} 
\end{equation}
\begin{equation}
\mathscr{E}^\text{(p)}_{b,n}(\mathcal{E}_z) = \frac{\gamma t}{4} n + \chg{\mathcal{O}( (\gamma t)^2 )}. \label{eq:process_Jz_qubits}
\end{equation}

{Note that (\ref{eq:process_Jz_qubits}) has been verified experimentally, for example by \citet{ozaeta_decoherence_2019}.}





\subsection*{\label{ssec:3:linear} Fit and deviation from the linear behaviour}

\begin{figure}[htbp!]
    \centering
    \includegraphics[width=\columnwidth]{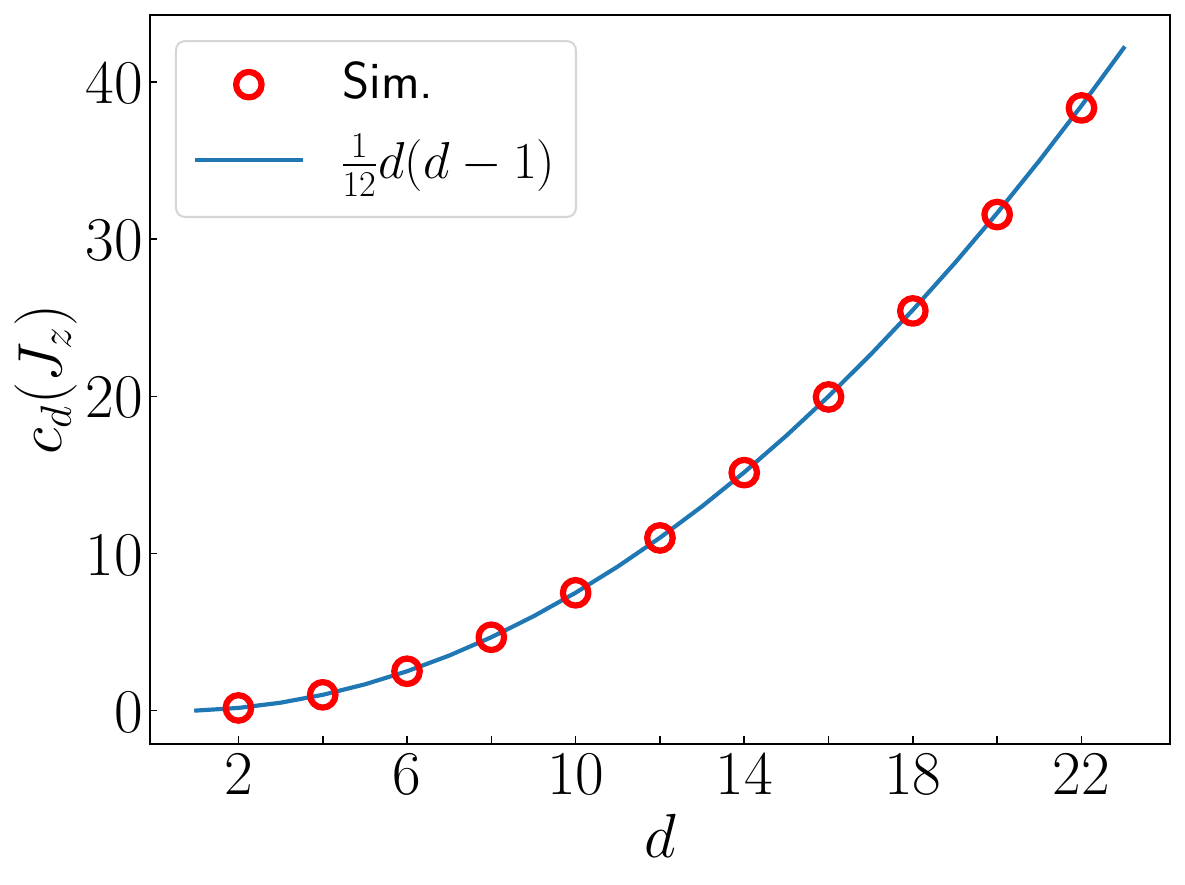}
    \caption{Rate of increase of $\overline{\mathscr{E}_d}(\mathcal{E}_z) = c_d(J_z)\gamma t$ as a function of qudit dimension for $H=\mathbb{0}_d$ and \notes{$\gamma t \in [0,10^{-4}]$}. The circled dots show the numerical results. The solid curve presents the expected {analytical} result {given by (\ref{eq:avEqudits}).}}
    \label{fig:1o12ddm1}
\end{figure}

\so{This procedure was then applied to} \chg{Using the procedures described in the \hyperref[sec:methods]{Methods section}, we simulated} single qudits {of dimension $d$} under pure dephasing, with $H=\mathbb{0}_d$ and small \notes{$\gamma t \in [0,10^{-4}]$} \notes{($\gamma \sim 10^{-4}$ in some \chg{nuclear spins in} molecular magnets such as in the experiments of \citet{godfrin_operating_2017})}. For example, the simulations were performed for even dimensions $d\in\llbracket2,22\rrbracket$. Fitting the AGIs $\overline{\mathscr{E}_d}(\mathcal{E}_z) = c_d\gamma t$ as a function of $\gamma t$ yielded the slopes $c_d$ that are shown in Fig.\ref{fig:1o12ddm1}, along with their {analytical} expression as a function of $d$ predicted in (\ref{eq:avEqudits}).

\begin{figure}[htbp!]
    \centering
    \includegraphics[width=\columnwidth]{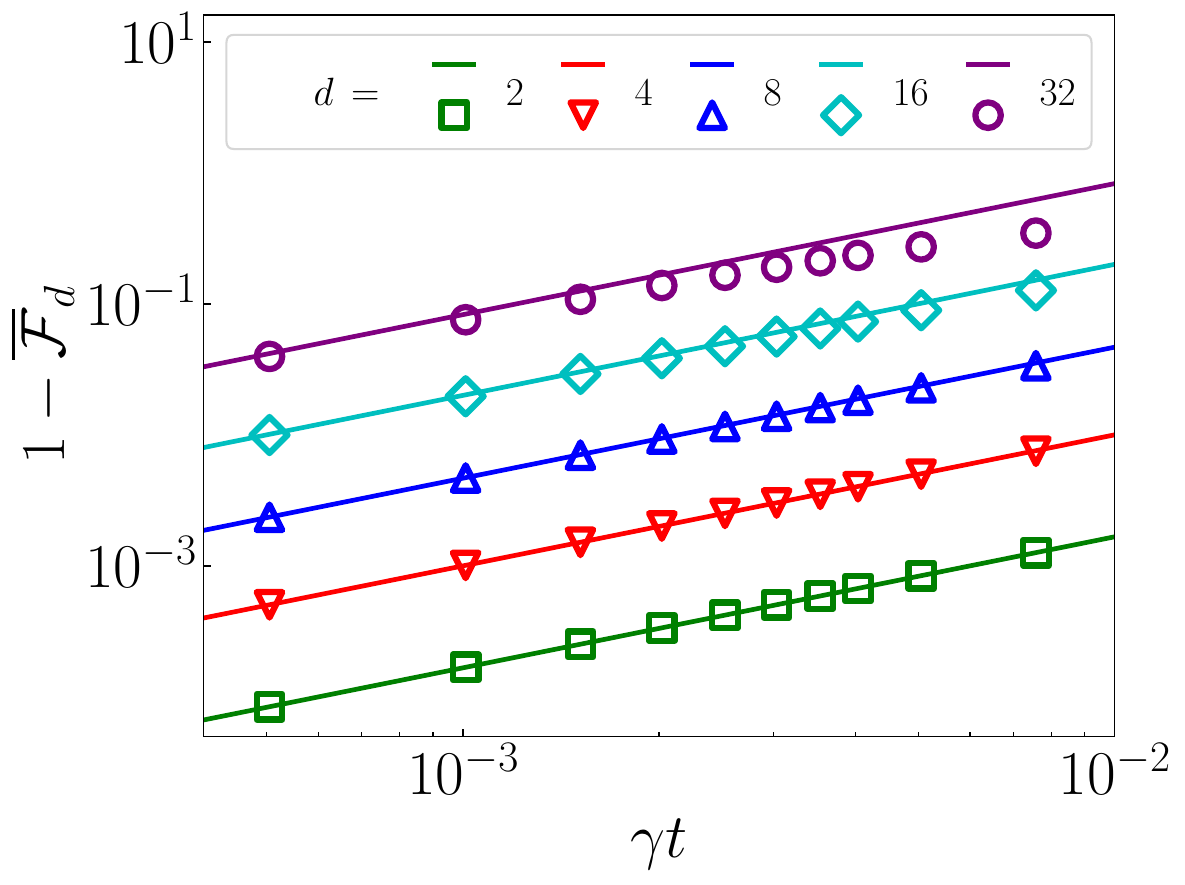}
    \caption{Average gate infidelities as a function of $\gamma t$. The data points show the computed values. {The solid lines represent} the linear theoretical behaviour \chg{from (\ref{eq:Fid112})}. Each colour/marker pair corresponds to a different value of $d$.}
    \label{fig:dev_lin}
\end{figure}

The same simulations were repeated for larger values of $\gamma t \in [5\times10^{-4},1\times10^{-2}]$ and $H=\mathbb{0}_d$. The AGIs were then computed and are shown in Fig.\ref{fig:dev_lin} alongside the linear infidelity predicted in (\ref{eq:avEqudits}). 

\begin{figure}[htbp!]
    \centering
    \includegraphics[width=\columnwidth]{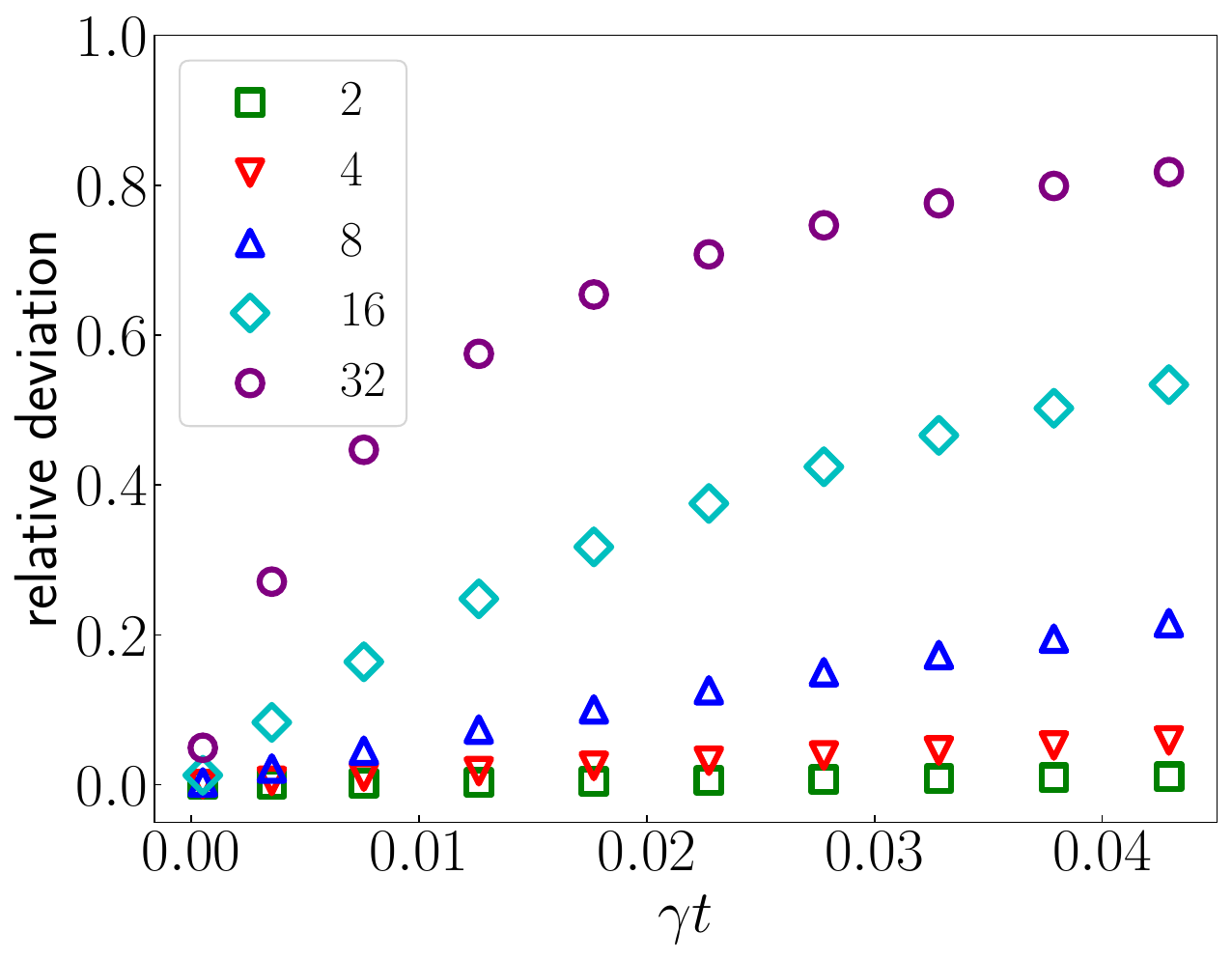}
    \caption{Relative deviation $1-\frac{\overline{\mathscr{E}_d}^\text{sim}}{\overline{\mathscr{E}_d}^\text{th}}$ as a function of $\gamma t$ for $H=\mathbb{0}_d$. $\overline{\mathscr{E}_d}^\text{sim}$ and $\overline{\mathscr{E}_d}^\text{th}$  were obtained from numerical computations and (\ref{eq:avEqudits}) {respectively}. Each marker corresponds to a different value of $d$.}
    \label{fig:dev_lin_quant}
\end{figure}

For more insight, Fig.\ref{fig:dev_lin_quant} shows the relative deviation of the computed infidelities from the expected first-order linear behaviour for a broader range of $\gamma t$ up to $5\times10^{-2}$.

Average gate infidelities linear in $\gamma t$ with gradients $\frac{d(d-1)}{12}$ were expected in the case of a single qudit under pure dephasing, according to (\ref{eq:avEqudits}). Fig.\ref{fig:1o12ddm1} supports this for small values of $\gamma t$: a least-squares fit of the computed gradients yields the expected relationship with \st{$1-R^2<10^{-8}$} \notes{$1-R^2<10^{-5}$}. Simulations for larger values of $\gamma t$ (Fig. \ref{fig:dev_lin} ) highlight deviations from this linear behaviour. These originate from $\chg{\mathcal{O}( (\gamma t)^2 )}$ terms of the form $(\gamma t)^{k>1}$ (see Eq. (\ref{eq:(gammat)k})). Moreover, for fixed values of $\gamma t$, as $d$ increases, the amplitude of this deviation is observed to increase (Fig.\ref{fig:dev_lin_quant}). This implies that the range of $\gamma t$ values for which the AGI can be treated linearly diminishes with increasing qudit dimension. Assuming a prefactor of the order $d^4$ for the $(\gamma t)^2$ term in the AGI series expansion (as \ref{eq:(gammat)2} hints), this provides an estimate of the range for which the deviation from linearity is negligible: $\gamma t \ll 1$ and $\dfrac{1}{d^2}$.

\subsection*{\label{ssec:3:gateindep} Gate dependence}

\begin{figure}[htbp!]
    \centering
    \includegraphics[width=\columnwidth]{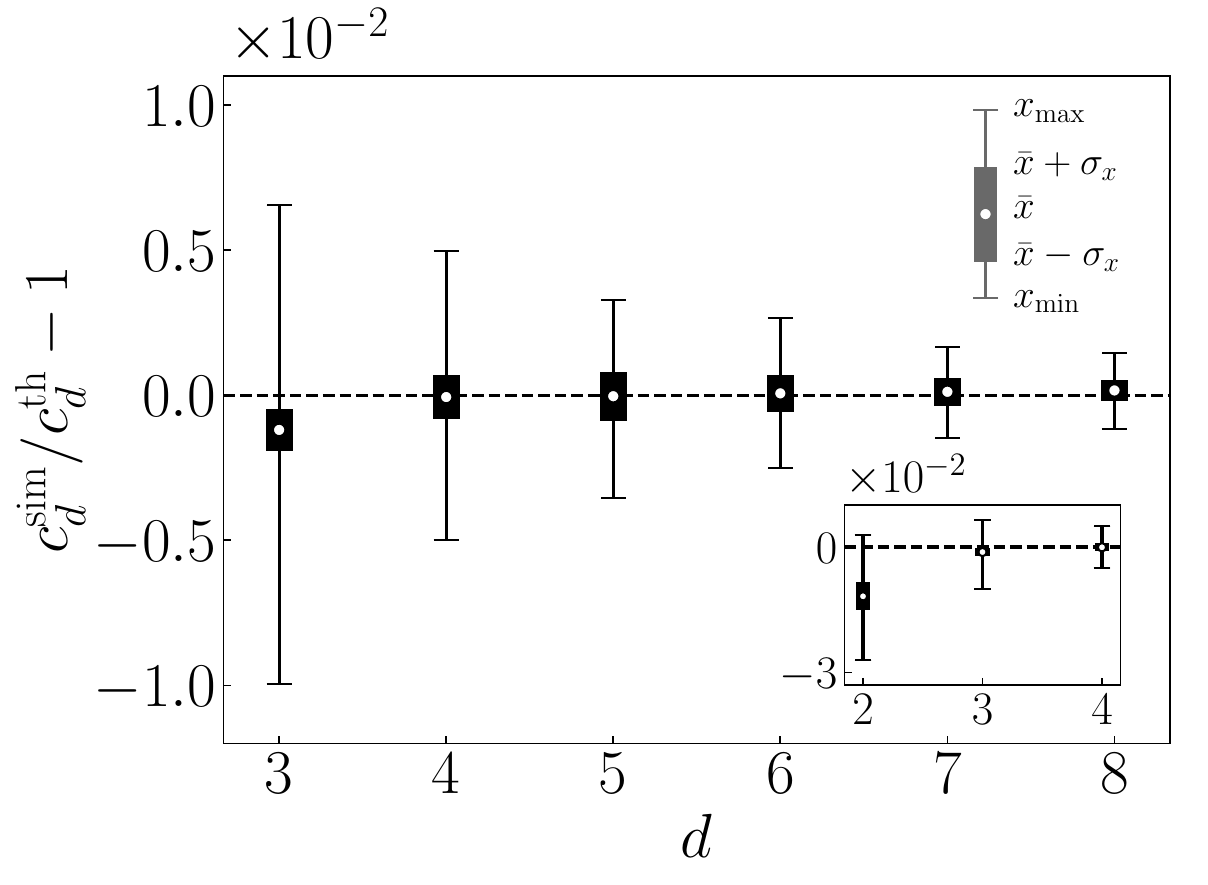}
    \caption{Statistical distributions of the relative deviation from the linear behaviour in (\ref{eq:avEqudits}) of the numerically obtained infidelity gradients $c_d$ for {$N_g=5000$} gates for $\gamma t \in [10^{-5},10^{-3}]$, as a function of the dimension $d \in \llbracket 3,8 \rrbracket$. The candlestick bar chart should be interpreted as indicated in the upper right\chg{, with $\sigma$ denoting the standard deviation}. The lower right inset shows the same results for $d \in \llbracket 2,4 \rrbracket$.}
    \label{fig:gate_dep}
\end{figure}

While the linearity of the AGI does not scale well with $d$, Eq. (\ref{eq:avEqudits}) {has another important characteristic that deserves to be studied}: the gate independence of the AGI. \notes{This was investigated over a large number of random gates for a given dimension $d$. Random unitary quantum gates in $U(d)$ were sampled from the circular unitary ensemble, which represents a uniform distribution over the unitary square matrices of dimension $d$, also known as the Haar measure on the unitary group $U(d)$,} and implemented on a qudit {through} a Hamiltonian obtained by gradient-ascent methods. \notes{We decided to model qudits as ladder systems, with one pulse per transition between adjacent levels as considered for example in the experiments of \citet{godfrin_operating_2017}, for a single-molecule magnet ($\text{TbPc}_2$, qudit with $d=4$), the $d-1$ pulses are then each represented by a control Hamiltonian in the interaction picture. \chg{More details are discussed in the \hyperref[sec:methods]{methods section}}. \chg{There are a large number of parameters that can influence the results under consideration, such as the free-evolution Hamiltonian or the matrix form of the control pulses, both of which are inherent to the physical realization. Therefore} other physical implementations and reference frames for the pulses can be considered, and the deviation from linearity they cause needs to be studied in more detail.} The AGIs were then computed for $\gamma t \in [10^{-5},10^{-3}]$ \notes{, which lie in the typical ranges observed in current platforms as seen in Table \ref{tab:platforms} \so{in subsection \ref{ssec:3:bvsd}}, } and their rate of increase as a function of $\gamma t$ was fitted. Fig.\ref{fig:gate_dep} shows the statistical distributions of the relative deviations {from the linear behaviour} of the obtained rates.

 Considering Fig.\ref{fig:gate_dep}, the relative deviation from the linear behaviour for different random gates seems to exceed 1\textperthousand $\,$ rarely and was not observed outside the $<$1\% range. Moreover, the range of deviations decreases as the dimension $d$ increases. The inset highlights a noticeable irregularity for $d=2$, a single qubit, where the relative deviation is of {the} order {of} 1\%. Note that for the $H=\mathbb{0}_d$ case simulated in Fig.\ref{fig:1o12ddm1}, this shift remained $<$1\textpertenthousand, coinciding with the dashed line in Fig.\ref{fig:gate_dep}, including {the case} $d=2$. Beginning at $d=2$, the gradient distributions appear broad and off-centred from the $H=\mathbb{0}_d$ case. As $d$ increases further, the distributions become progressively concentrated around $0$. The gate-dependence also arises from $\chg{\mathcal{O}( (\gamma t)^2 )}$ terms, with $\gamma t^2$ being dominant in the $\gamma t^2 ||H|| \gtrsim 1 $ regime {(see Eq. (\ref{eq:gammat2}))}. Therefore, the AGI can only be considered gate-independent when $\gamma t \ll 1$ and $\gamma t \ll \dfrac{1}{||H||t}$. An informative figure showing the deviation from linearity and gate-dependence at higher values of $\gamma t$ is available in Appendix \ref{apdx:devlin}.

\subsection*{\label{ssec:3:Jxy} Other cases than pure dephasing}

\begin{figure}[htbp!]
    \centering
    \includegraphics[width=\columnwidth]{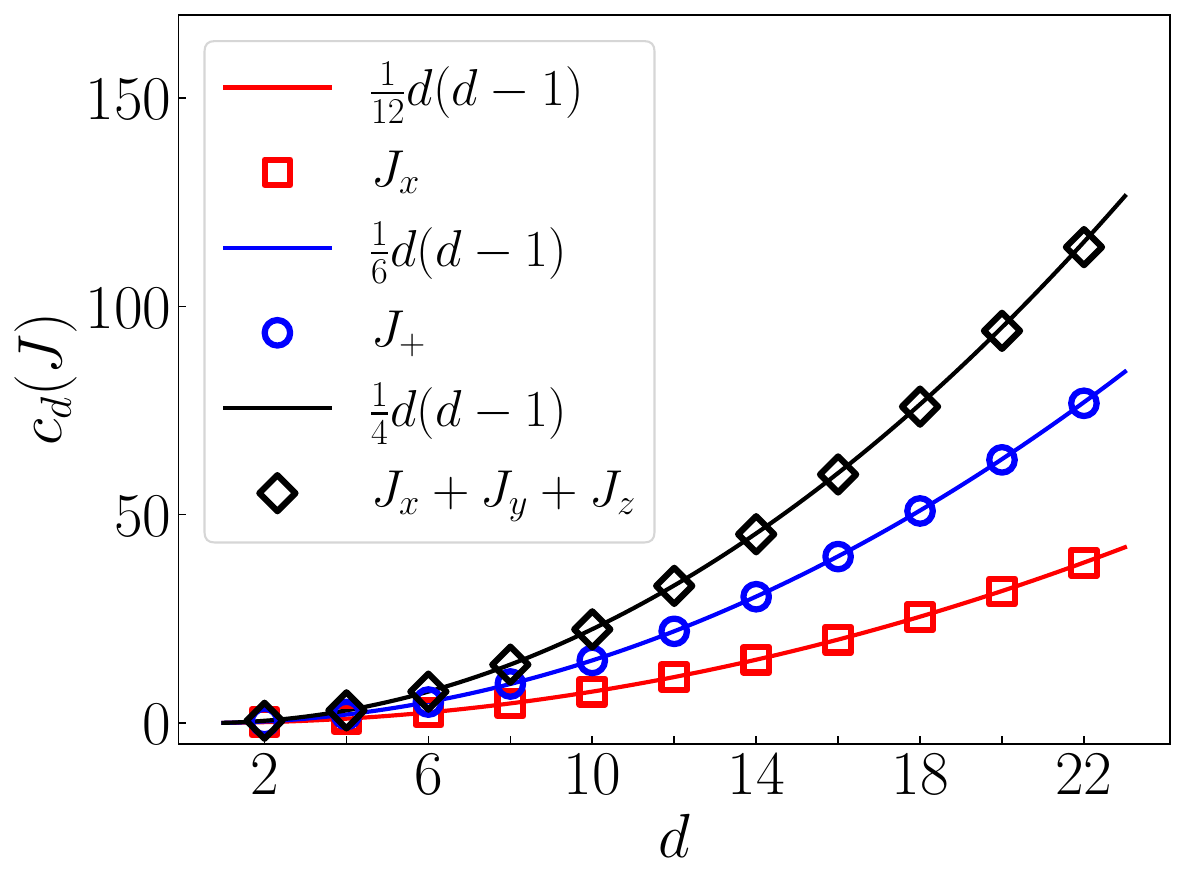}
    \caption{Rate of increase of $\overline{\mathscr{E}_d}(\mathcal{X}) = c_d(J)\gamma t$ as a function of {the} qudit dimension with $H=\mathbb{0}_d$ and \notes{$\gamma t \in [0,10^{-4}]$} . The markers show the numerical results. The solid curves represent the expected linear responses \chg{according to (\ref{eq:avEquditsLsimp})}. Each marker/colour pair corresponds to a different error channel $\mathcal{X}$, with collapse operators $J$ specified in the legend.}
    \label{fig:diffLs}
\end{figure}

Fig.\ref{fig:diffLs} shows AGI rates of increase for channels different from pure dephasing {namely}: $\overline{\mathscr{E}_d}(\mathcal{E}_x)$, $\overline{\mathscr{E}_d}(\mathcal{E}_+)$, {and} $\overline{\mathscr{E}_d}(\mathcal{E}_{x,y,z})$ corresponding to bit-flip, amplitude damping and depolarizing channels respectively. The simulations were performed again with $H=\mathbb{0}_d$, small $\gamma t \in [0,10^{-4}]$ and for even dimensions $d\in\llbracket2,22\rrbracket$.

Consider {${\cal R}$}, the unitary transformation representing a change of basis, such as a $3D$ real-space rotation. The average gate fidelity defined in (\ref{eq:avgatefid}) is invariant under the transformation {$\rho \rightarrow {\cal R}^\dag \rho {\cal R}$}. This is supported by a comparison of the results for $L=J_z$ and $L=J_x$ in Fig.\ref{fig:1o12ddm1} and Fig.\ref{fig:diffLs}, respectively, since the two gradients appear to share the same dependency in $d$. Moreover, let $\{l_k\}$ be an ensemble of traceless collapse operators with corresponding error channels $\{e_k\}$, then define  $L=\sum_k l_k$ and associated error channel $\mathcal{E}$. From (\ref{eq:krausfid}) and (\ref{eq:krausop}), as long as $\Tr\left(l_k^\dag l_j\right)=0,\,\forall j\neq k$ then $\overline{\mathscr{E}}(\mathcal{E}) = \sum_k \overline{\mathscr{E}}(e_k)$. Fig. \ref{fig:diffLs} again supports such behaviour since simulations with the collapse operator $L = J_+ {\equiv} J_x + i J_y$ yield gradients twice as large as, and $L=J_x+J_y+J_z$ yield gradients three times as large as, the $L=J_z$ case.

\subsection*{\label{ssec:3:bvsd}{A single qudit vs an ensemble of qubits}}

\begin{figure}[htbp!]
    \centering
    \includegraphics[width=\columnwidth]{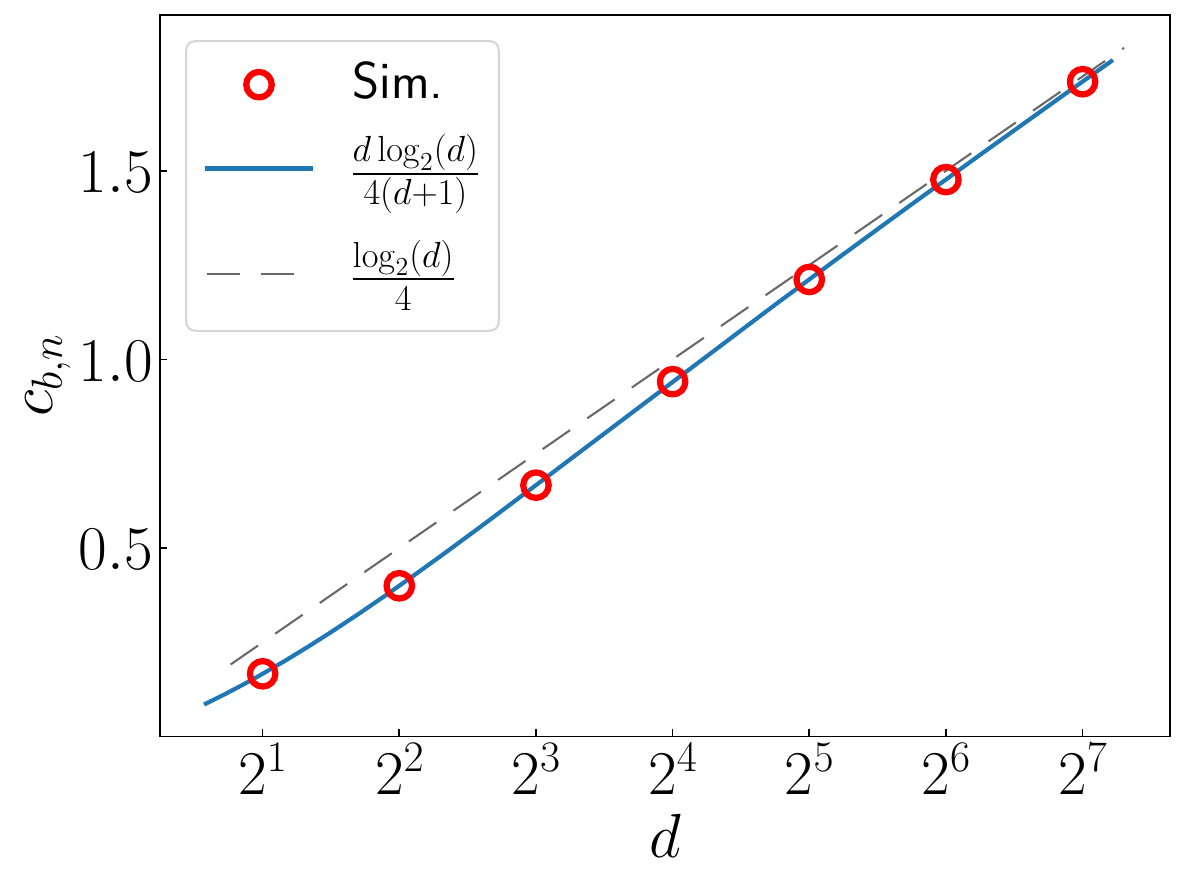}
    \caption{Rate of increase of $\overline{\mathscr{E}_{b,n}}(\mathcal{E}_z) = c_{b,n}(\chg{\{L_k\}})\gamma t$ as a function of qudit dimension $d=2^n$ with $H=\mathbb{0}_{2^n}$ and \notes{$\gamma t \in [0,10^{-4}]$}. \chg{The $\{L_k\}$ collapse operators are the ones defined in (\ref{eq:Lkqubit}).} The circled dots show the numerical results. The solid curve presents the expected theoretical result \chg{according to (\ref{eq:avEqubits})}. The dashed line shows $\mathscr{E}^\text{(p)}_{b,n}(\mathcal{E}_z) $ given in (\ref{eq:process_Jz_qubits}) which is linear in $n={\mathrm{log}}_2(d)$.}
    \label{fig:qubits}
\end{figure}

An ensemble of $n$ qubits were simulated under identical pure dephasing, with $H=\mathbb{0}_{2^n}$ and small \notes{$\gamma t \in [0,10^{-4}]$}. The simulations were performed for $n\in\llbracket1,7\rrbracket$. Fitting the AGI $\overline{{\mathscr{E}_{b,n}}(\mathcal{E}_z)} = c_{b,n}\gamma t$  as a function of $\gamma t$ yielded the slopes $c_{b,n}$ that are shown in Fig.\ref{fig:qubits}, along with their {analytical} expression as a function of $d$ {given} in (\ref{eq:avEqubits}).

\begin{figure}[htbp!]
    \centering
    \includegraphics[width=\columnwidth]{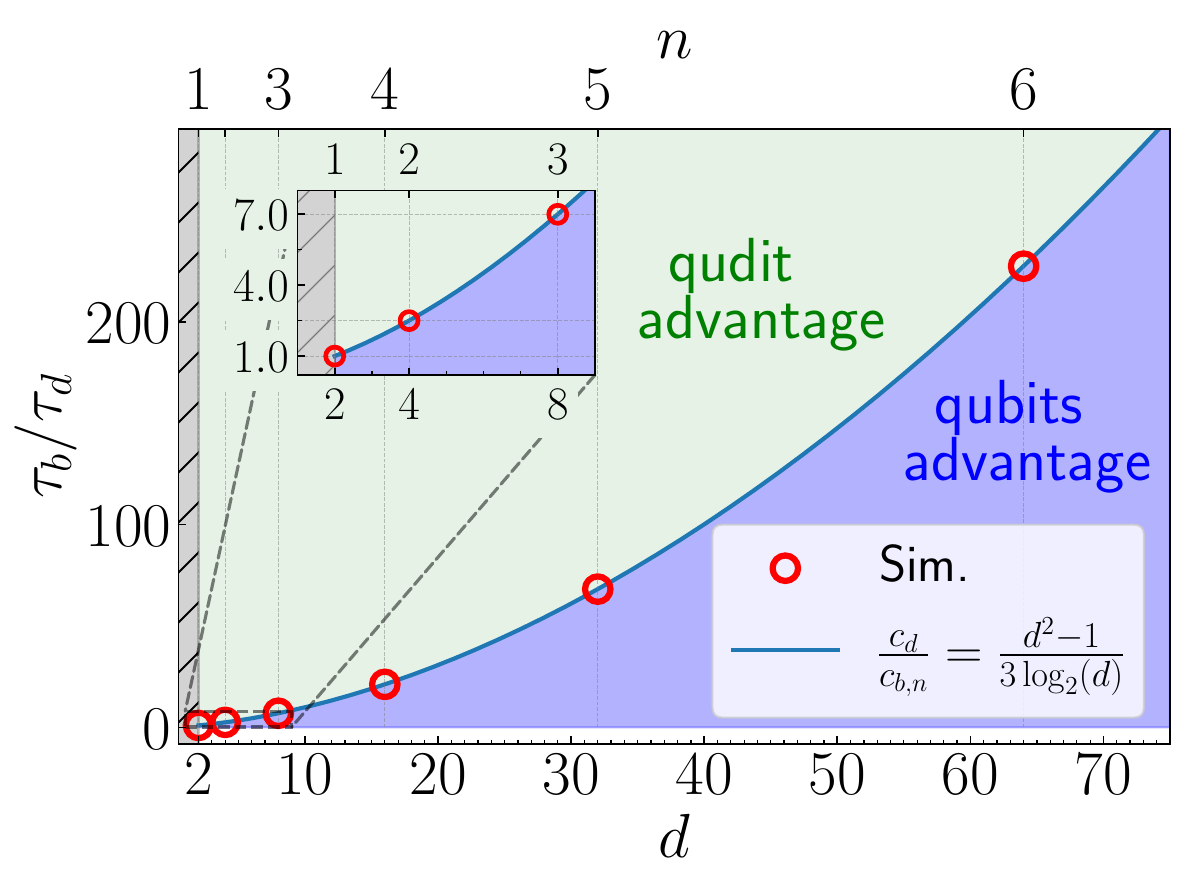}
    \caption{Potential range for $\tau_{b} / \tau_{d}$. The rounded circles show the numerical values obtained for $c_{d}/c_{b,n}$. The solid curve comes from (\ref{eq:conditionT2}) and highlights the theoretical critical values of ${T_{2,d}}/{T_{2,b}}$.}
    \label{fig:bvd}
\end{figure}

The same simulations were performed {on a single qudit} with {dimension} $d=2^n$ and Fig.\ref{fig:bvd} shows the ratios $\frac{c_d}{c_{b,n}}$ for $n\in\llbracket1,6\rrbracket$ as well as the theoretical curve {provided by Eq. (\ref{eq:conditionT2})} on which the points should be falling. According to the same {Eq. (\ref{eq:conditionT2})}, this curve also highlights the critical values of $ \tau_{b} / \tau_{d} $, denoting the figure of merit $\tau_k = t_k/T_{2,k} = \gamma_k t_k /2$, with respect to qudit/qubits advantage in terms of the rate of increase of the AGI.

The AGI gradients obtained for an ensemble of $n$ qubits under identical pure dephasing were expected to follow a $\frac{d\log_2(d)}{4(d+1)}$ relationship as a function of $d$ (\ref{eq:avEqubits}). Fig.\ref{fig:qubits} justifies this for small values of $\gamma t$ with the least-square fit now yielding \st{$1-R^2<10^{-10}$} \notes{$1-R^2<10^{-7}$}. Finally, Fig. \ref{fig:bvd} provides quantitative data for the ratio of the decoherence times of {a} single qudit vs an ensemble of qubits. Some values of interest are summarised in Table \ref{tab:specdvb}. For example, in order for a \notes{qu-$8$-it {(qudit with $d=8$)} } to present a computational fidelity advantage over $3$ qubits for a fixed gate time, the qudit platform needs a coherence time at least $7$ times longer than the multiqubit platform. \new{Note that an intuitive scaling such as $\frac{d^2}{\log_2(d)}$ would indicate a much more demanding constraint of $21.5$.}

\bgroup
\def\arraystretch{2}
\setlength\tabcolsep{5pt}
\begin{table}[htbp!]
\begin{tabular}{|l||c|c|c|c|}
\hline
\textbf{Number $n$ of qubits}             & $1$ & $2$   & $3$ & $6$      \\ \hline
\textbf{Dimension $d$ of the qudit}        & $2$ & $4$   & $8$ & $64$    \\ \hline
\textbf{Critical $\tau_{b} / \tau_{d}$} & $1$ & $2.5$ & $7$ & $227.5$ \\ \hline
\end{tabular}
\caption{Ratios of gate times in units of decoherence times between qubits and qudits for specific values of $n$ and $d$. $\tau_{b} / \tau_{d}$ needs to be larger than the critical values in order for a single qudit to be advantageous vs an equivalent ensemble of $n$ qubits.}
\label{tab:specdvb}
\end{table}
\egroup

\bgroup
\def\arraystretch{2}
\setlength\tabcolsep{0mm}

\begin{table}[htpb!]
\begin{tabular}{cc<{\hspace{12pt}}c<{\hspace{12pt}}r<{\hspace{12pt}}r<{\hspace{12pt}}c<{\hspace{12pt}}l}

 & $\;\,d$ & $\;\,n$ & \multicolumn{1}{c}{$T_{2}$}  & \multicolumn{1}{c}{$t_{n}$}     & $\tau_{n}$   & ref. \\ \hline
\rowcolor{blue!75!gray!10} 
\cellcolor{white} & $\;2$ & $\;2$ & $\sim$ 10 µs & 60 ns & $\sim 10^{-2}$ & \citex{kjaergaard_programming_2020} \\
\rowcolor{green!75!gray!15} \cellcolor{white}& $\;2$ & $\;2$ & $\sim 2$ µs  & $51$ ns & $\sim 10^{-2}$ & \citex{madjarov_high-fidelity_2020}  \\ 
\rowcolor{blue!75!gray!10} \cellcolor{white}& $\;2$ & $\;17$ & $\sim$ 30 µs & $\sim$ 100 ns & $\sim 10^{-3}$ & \citex{krinner_realizing_2022} \\ 
\rowcolor{orange!15} \cellcolor{white}  & $\;2$ & $\;24$ & $\sim 100$ ms  & $\sim 200$ µs & $\sim 10^{-3}$ & \citex{pogorelov_compact_2021}\\
\rowcolor{cyan!15} \cellcolor{white}\multirow{-5}{.05\columnwidth}{\rotatebox[origin=c]{90}{qubits}} & \chg{$\;2$} & \chg{$\;1$} & \chg{$\sim 1$ ms}  & \chg{$\sim 1$ µs} & \chg{$\sim 10^{-3}$} & \chg{\citex{PhysRevApplied.19.064060}}\\  \hline  
\rowcolor{Magenta!15} \cellcolor{white}& $\;4$ & $\;1$           & 0.32 ms     & $\sim$ 100 ns & $\sim 10^{-4}$ &  \citex{moreno-pineda_molecular_2018}\\ 
\rowcolor{blue!75!gray!10} 
\cellcolor{white} & $\;4$ & $\;1$ & $\sim 100$ µs & $\sim 150$ ns & $\sim 10^{-3}$ & \cite{PhysRevLett.125.170502,cao2023emulating} ('20,'23)\\
\rowcolor{orange!15} \cellcolor{white}& $\;3$ & $\;2$ & $\sim 100$ ms  & $\sim 100$ µs & $\sim 10^{-3}$ & \citex{ringbauer_universal_2022}    \\  
\rowcolor{gray!15} \cellcolor{white}& $\;4$ & $\;2$& $\infty$\footnote{considered unlimited by the source authors}  & \multicolumn{1}{c}{--} & 0 & \citex{chi_programmable_2022} \\ 
\rowcolor{green!75!gray!15} \cellcolor{white}\multirow{-4}{.05\columnwidth}{\rotatebox[origin=c]{90}{qudits}}& $\;52$ & $\;1$  & \multicolumn{1}{c}{--}  & $\sim 100$ ns\footnote{no universal gates for the moment, only specific quantum operations implemented} & \multicolumn{1}{l}{$\quad\;$--} & \citex{larrouy_fast_2020} 
\\ \hline
\end{tabular}
\bgroup\scriptsize
\\\colorbox{green!75!gray!15}{Rydberg atoms},
\colorbox{orange!15}{trapped ions},
\colorbox{cyan!15}{\chg{electronic spins in molecular magnets}}\\
\colorbox{blue!75!gray!10}{superconducting qubits},
\colorbox{Magenta!15}{\so{molecular} nuclear spins \chg{in molecular magnets}},
\colorbox{gray!15}{photonic qudits}
\egroup
\caption{Decoherence times ($T_2$) and gate times ($t_n$) of different qubit/qudit platforms. $d$ and $n$ are the maximum dimension and number of qudits an operation was applied to, while $\tau_n=t_n/T_2$ is the figure of merit.}
\label{tab:platforms}
\end{table}
\egroup

From Table \ref{tab:platforms}, state-of-the-art single qudit platforms, such as trapped ions \cite{ringbauer_universal_2022}, present coherence times of the order of $100$ms for a single qu-$7$-it, orders of magnitude longer than superconducting qubits \cite{ozaeta_decoherence_2019, chen_qubit_2014, kjaergaard_superconducting_2020, rosenblum_cnot_2018}. Trapped ions present $\gamma t\approx 10^{-3}$, while $\gamma t\approx 10^{-2}$ for superconducting qubits; this ratio of 10 would allow qudits with $d\lesssim 10$ to still be advantageous \nnotes{i.e., according to (\ref{eq:conditionT2}), the single qu-$7$-it would still maintain a higher average gate fidelity over one gate acting on the whole Hilbert Space than the multiqubit platform}. Another comparison with superconducting qubits could be molecular nuclear \chg{spin} qudits, where some proposals put $\gamma t\approx 10^{-4}$ \chg{(see \citet{moreno-pineda_molecular_2018})}, and whose coherence times are $\sim6-7$ times larger than the superconducting qubit case. With figure of merits $\tau$ $\sim 100$ larger than superconduction platforms, single molecular nuclear \chg{spin} qudits with $d\lesssim 40$ are still advantageous over equivalent superconducting qubits, i.e. $n\sim 5$. Such high-$d$ qudit platforms can still be conceivable, given that some specific quantum operations on $d=52$ have already successfully been implemented on, for example, Rydberg atoms \cite{larrouy_fast_2020}. \notes{However, it remains to be seen if universal quantum gate generation will become easily achievable in practice with such high $d$.}

Finally, one can compare (\ref{eq:conditionT2}) and (\ref{eq:conditionT2d^N}) to discuss conditions on $N$ qudits outperforming $N\log_2(d)$ qubits. From this, if a single qudit outperforms $\log_2(d)$ qubits, the advantage remains conserved as long as the multiqudit gate time scales slower from $1$ qudit to $N$ qudits than the multiqubit from $\log_2(d)$ to $N\log_2(d)$ qubits.





\subsection*{\label{sec:concl} Conclusion}
{Given the rapid development of quantum computing platforms with very different physical properties, such as decoherence time or Hilbert space dimension (see Table II), there is a growing need for detailed elaboration of the tradeoffs between their information density and noise error rates. By combining analytical results and numerical simulations, we have performed a comparative study of gate efficiency for systems composed of sets of qubits or qudits.}   
A fluctuation-dissipation-like relation for the gate infidelity of an operation on a pure state was derived. {We} then put forward a physically-informed method to obtain the first-order effect of Markovian noise on the \textit{average gate infidelity} (AGI). A connection was made between the latter and the first gate-independent result. The rate of increase of the AGI of a single qudit vs equivalent multiple qubits under pure dephasing was compared. This yielded a critical curve of the ratio of their respective gate times in units of decoherence time, a quantity indicating how time-efficient operations on a particular system are. Values on either side of the curve specify which of the two systems had a higher rate of increase of the AGI. To compete in terms of gate fidelity, as the dimension increases, the efficiency of qudit gates must \new{not simply} always be larger than the multiqubit one by a factor $O(d^2/\log_2(d))$\new{, but precisely by a factor $\frac{d^2-1}{3\log_2(d)}$, which makes a significant difference for lower values of $d$ for which it provides less demanding constraints}. Additionally, {analytical} expressions of linear response for arbitrary collapse operators and a general multiqudit system were presented {(see (\ref{eq:avEquditsLsimp}) and (\ref{eq:d^Ngene}))}. {They may be useful to those working in the field of quantum computing \nnotes{e.g. to, as mentioned in \so{Sec.\ref{sec:intro}} \chg{the introduction}, benchmark qudit platforms either in terms of maximal practical $d$ or in terms of conditions on the figure of merit to compensate for the greater noise scaling, in comparison with current state-of-the-art multiqubit platforms}.}

Numerical simulations contributed to the discussion on the validity and limits of the linear response assumption. This further restricted the ranges of possible $\gamma t \ll 1$ accounting for qudit dimension, gate, and noise type. For example, the larger the dimension, the lower the relative gate-dependent response. Finally, after simulations supported the analytical critical curve, different current platforms were studied with respect to this condition on gate time efficiency. Given equivalent Hilbert {space} dimensions, viable qudit platforms \nnotes{\chg{(leveraging advantageous decoherence times and gate speeds to compensate for the higher rate of increase in AGI)}} capable of outperforming equivalent state-of-the-art multiqubit ones in gate fidelity have been found for pure dephasing. Moreover, this performance could be extended to qudits with $d$ as large as $\sim 40$ in the case of nuclear \chg{spins in} molecular magnets, for example. Some multiqubit platforms still outperform any existing qudit platform regarding scalability in the number of subsystems. However, it is conceivable that some scalable qudit platforms continue to outperform equivalent multiqubit systems in terms of attainable fidelity. Further study of how multiqudit and multiqubit gate times scale with the number of subsystems is needed. {Moreover, this study was limited to first-order noise responses. However, \notes{using the notation qu-$j$-it for a qudit of dimension $d=j$}, \notes{through carefully chosen quantum error correction schemes, it was} recently demonstrated it is possible to entirely remove the first-order response of logical \notes{qu-$k$-its} embedded in physical \notes{qu-$d$-its} ($k<d$) through carefully chosen encodings \cite{chiesa_embedded_2021, petiziol_counteracting_2021}.  \chg{Of particular interest in the authors' future work is the study of the Hamiltonian-dependent response of the dimension-dependent AGI. More generally,} an additional study of higher-order responses of logical qudits vs. physical qubits would therefore also be required to assess the viability of these logical error-resilient qudits.} This could elucidate if \nnotes{the presence of single} qudit advantage\nnotes{, as quantified by this paper,} is robust to system scaling and if qudits will remain useful beyond the NISQ era.



{\color{redd}
\section*{Methods}

\subsection*{Numerical noisy qudit/multiqubit simulation}
All simulations were done using the \texttt{Python} package \texttt{QuTiP} \cite{johansson_qutip_2012} version 4.7, \texttt{SciPy} version 1.7.3, and \texttt{NumPy} version 1.21.5. This subsection aims to present the \textit{modus operandi} for obtaining the numerical results referenced in the different figures: the AGIs ($\overline{\mathscr{E}})$ for Fig.\ref{fig:dev_lin} and Fig.\ref{fig:dev_lin_quant} and the slope of the AGIs ($c$) in the other figures.\\

Standard packages: Essential functions for simulating quantum dynamics and fitting curves to data are provided by the \texttt{QuTiP} library, including functions for propagator calculation in superoperator form (\texttt{qt.propagator}) and gate fidelity evaluation (\texttt{average\_gate\_fidelity} from \texttt{qutip.metrics}), along with the \texttt{curve\_fit} function from \texttt{scipy.optimize}.\\
    
Parameters: We define the system's dimension $d$, the decay parameter $\gamma$, and the collapse operators $\{L_k\}$ under consideration. The collapse operators are \texttt{QObj} instances characterized by their matrix form in the canonical basis. Additionally, we generate a list of time points for simulating the system's evolution. Considering the quantity of interest in this study is $\gamma t$, $\gamma$ is chosen as fixed, and the range of $\gamma t$ is then given by the range of the time points.\\

Time evolution: The simulation of the quantum system's time evolution is facilitated by computing the propagator using the system's Hamiltonian, the list of time points, and the collapse operators multiplied by $\sqrt{\gamma}$. This generates a time-dependent propagator in the form of a list of $\texttt{QObj}$ superoperators for different values of $\gamma t$. The system's Hamiltonian will be discussed in further detail in the following subsection. However, apart from Fig.\ref{fig:gate_dep} studying the gate/Hamiltonian-dependence, the other figures report simulations done with a vanishing Hamiltonian $H=\mathbb{0}_d$ since the quantities under consideration are considered Hamiltonian-independent.\\

Fidelity Calculation: At each $\gamma t$ the average gate fidelity is computed relative to a target gate, in the case of $H=\mathbb{0}_d$: the identity matrix. This is the quantity displayed in Fig.\ref{fig:dev_lin} and Fig.\ref{fig:dev_lin_quant}.\\

Curve Fitting: A curve is fitted to the calculated fidelities over the range of $\gamma t$ using the \texttt{curve\_fit} function. This process involves fitting the function $1-c\gamma t$ for the parameter $c$. The obtained slopes $c(\{L_k\})$ are then the ones displayed in the figures \ref{fig:1o12ddm1}, \ref{fig:gate_dep}, \ref{fig:diffLs}, \ref{fig:qubits} and \ref{fig:bvd}. Moreover the least-square fit parameter $R$ given by the fitting functions is the one reported in this study.

\subsection*{Random gate and pulse Hamiltonian generation\label{sec:methods}}

Gate generation: In the \hyperref[ssec:3:gateindep]{study of the gate-dependent deviation from the analytical results of this manuscript}, for each dimension $d$ under consideration, a set of $N_g$ = 5000 gates have been randomly generated with the \texttt{Bristol}\cite{suezen_bristol_2017} package in \texttt{Python}. The gates have been drawn from the circular unitary ensemble, and are thus considered to be uniformly distributed over the Haar measure. Subsequently, to generate an associated set of pulses for each gate, we have used the \texttt{optimize\_pulse\_unitary} function from the pulse optimization module (\texttt{control.pulse\_optim}) of \texttt{QuTiP}. \\

Pulse generation: The pulse generation is done through gradient-ascent methods using the \texttt{GRAPE} algorithm \cite{khaneja_optimal_2005} and was run in parallel for each gate using a high-performance cluster. The numerical optimizer used by default is the L-BFGS-B method. Assuming the control hamiltonian, as discussed after (\ref{eq:master}), takes the form

\begin{equation}
    H_c(t) = \sum_{k=0}^N u_k(t) H_k,
\end{equation}

with $H_k$ being a basis set of controls and $u_k(t)$ representing the time-dependent control amplitudes, the optimization process involves finding the set of $u_k(t)$ that best approximates the target gate. \\

Choice of Hamiltonian: For the simulations reported in this paper, we decided to model qudits as ladder systems, with one pulse per transition between adjacent levels as considered for example in the experiments of \citet{godfrin_operating_2017}, for a single-molecule magnet ($\text{TbPc}_2$, qudit with $d=4$), the $d-1$ pulses are then each represented by two control Hamiltonians in the interaction picture. More explicitly, the basis set of controls is chosen to be the ensemble of pairs $\ket{k}\bra{k+1} + \ket{k+1}\bra{k}$ and $i(\ket{k}\bra{k+1} - \ket{k+1}\bra{k})$, with $k$ running from 1 to $d-1$. Moreover, $H_0$, the free-evolution, is chosen to be vanishing since we consider the interaction reference frame.
    
}


\section*{Data Availability statement}

The numerical data presented in this study are available from the authors upon
reasonable request.

\section*{Code Availability statement}

The code for numerical simulations is available from the authors upon reasonable request.


\begin{acknowledgments}
This work was funded by the French National Research Agency (ANR) through the Programme d'Investissement d'Avenir under contract ANR-11-LABX-0058\_NIE and ANR-17-EURE-0024 within the Investissement d’Avenir program ANR-10-IDEX-0002-02. D.J. and M.R. gratefully acknowledge financial support from the Deutsche Forschungsgemeinschaft (DFG, German Research Foundation) through the Collaborative Research Centre “4f for Future” (CRC 1573, project number 471424360) project B3. J-G.H. also acknowledges QUSTEC funding from the European Union’s Horizon 2020 research and innovation program under the Marie Skłodowska-Curie Grant Agreement No. 847471. The authors would like to acknowledge the High Performance Computing Center of the University of Strasbourg for supporting this work by providing scientific support and access to computing resources. Part of the computing resources were funded by the Equipex Equip@Meso project (Programme Investissements d'Avenir) and the CPER Alsacalcul/Big Data.

\end{acknowledgments}


\section*{Author Contributions}

D.J. conceptualized the study, derived analytical results, and drafted the manuscript with subsequent input of all the authors. D.J. and J-G.H. designed the methodology of the study with the input of P-A.H. . D.J and J-G.H. wrote the computational code and performed numerical simulations. D.J. generated and prepared the figures. P-A.H and M.R. supervised and conceptualized the project and advised on the manuscript's drafting. All authors participated in the proofreading and the preparation of the manuscript.

\section*{Competing Interests}

The authors declare no competing interests.

\section*{References}

\appendix

\section{Complementary Derivations}

\subsection{Fluctuation-dissipation relation} \label{apdx:flucdissip}

Substituting Eq.(\ref{eq:rhopertresult}) and Eq.(\ref{eq:rhotpert}) into Eq.(\ref{eq:deffidsimpl}) and Eq.(\ref{eq:deferr}) leads to
\begin{equation}
    \mathscr{E}(\rho^*) = \gamma t \left( \frac{1}{2} \Tr\left(\rho^* \left\{ L^\dag L, \rho^* \right\}\right) - \Tr\left(\rho^* L \rho^* L^\dag   \right)\right) + \chg{\mathcal{O}( (\gamma t)^2 )}. 
\end{equation}

The trace being invariant by cyclic permutations and ${\rho^*}^2 = \rho^*$ leads to the simplication
\begin{equation}
    \frac{1}{2} \Tr\left(\rho^* \left\{ L^\dag L, \rho^* \right\}\right) = \Tr\left(\rho^*L^\dag L\right) \equiv \langle L^\dag L\rangle_{*}\;.
\end{equation}

Moreover
\begin{align}
    \Tr\left(\rho^* L \rho^* L^\dag   \right) &= \Tr\left( \ket{\varphi^*}\ev{L}{\varphi^*}\bra{\varphi^*}L^\dag\right) \\ &= \ev{L}_* \Tr\left( \rho^*L^\dag \right) \\ &= \langle L^\dag \rangle_*\ev{L}_* \;.
\end{align}

{Accounting for the above results, one finally obtains}
\begin{equation}
    \mathscr{E}(\rho^*) = \gamma t \left( \langle L^\dag L\rangle_{*} - \langle L^\dag \rangle_*\ev{L}_* \right) + \chg{\mathcal{O}( (\gamma t)^2 )}, 
\end{equation}
which can be rewritten as Eq. (\ref{eq:flucdissip}).

\subsection{Average Gate Infidelity for the Pure Dephasing Channel of one qudit}\label{apdx:puredephkraus}

{In a trivial way we have} $\Tr(E_1) \propto \Tr(J_z) = 0$, and
\begin{align*}
    &\Tr(E_0) = d  - \frac{\gamma t}{2} \Tr(J_z^2) \\ &= d - \frac{\gamma t}{2} \sum_{k=0}^{d-1} \left(\frac{d-1-2k}{2}\right)^2 \\
    &= d - \frac{\gamma t}{8} \left[  d(d-1)^2 - 4 (d-1) \sum_{k=0}^{d-1} k + 4 \sum_{k=0}^{d-1} k^2\right]\\
    &= d - \frac{\gamma t}{8} \left[  d(d-1)^2 - 2 d (d-1)^2 + 4 \frac{d(d-1)(2d-1)}{6}\right]\\
    &= d - \frac{\gamma t}{24} d(d^2-1),
\end{align*}
{which results in}
\begin{equation}
    \left|\Tr(E_0)\right|^2 = d^2 - \frac{\gamma t}{12} d^2(d^2-1) + \chg{\mathcal{O}( (\gamma t)^2 )} \;. 
\end{equation}

Therefore
\begin{equation}
\label{eq:Error112}
    \overline{\mathcal{F}}(\mathcal{E}_z) = \frac{d + d^2 - \frac{\gamma t}{12} d^2(d^2-1)}{d(d+1)} + \chg{\mathcal{O}( (\gamma t)^2 )} = 1 - \frac{\gamma t}{12} d(d-1)  + \chg{\mathcal{O}( (\gamma t)^2 )} \;, 
\end{equation}
{which leads to the simplified expression (\ref{eq:Fid112}).}

\chg{Moreover, this is the origin of the factor $\frac{1}{12}$ in (\ref{eq:avEqudits}) whose ratio with the $\frac{1}{4}$ obtained in (\ref{eq:avEqubits}) and (\ref{eq:TrE0qubits}) leads to the non-trivial factor $\frac{1}{3}$ in (\ref{eq:conditionT2}).}

\subsection{Average Gate Infidelity for the pure dephasing channel of $n$ qubits} \label{apdx:agiqubits}

{Since} $\Tr(E_k) = 0 \; \forall k \ne 0$, only $\Tr(E_0)$ is left and {is given by}
\begin{equation}
\begin{aligned}
    \Tr(E_0) &= 2^n  - \frac{\gamma t}{2} \sum_{k=1}^n \left[\Tr(\left.S^2_{z}\right.^{(k)}) \prod_{j\neq k} \Tr(\left.\mathbb{1}_2^2\right.^{(j)})\right] \\
    &= 2^n  - n \frac{\gamma t}{8} 2^n,
\end{aligned}
\end{equation}
 {leading to}
\begin{equation}
\label{eq:TrE0qubits}
    \left|\Tr(E_0)\right|^2 = 2^{2n} - \frac{\gamma t}{4} n2^{2n} + \chg{\mathcal{O}( (\gamma t)^2 )}\;, 
\end{equation}
{which allows to obtain} (\ref{eq:avEqubits}) using (\ref{eq:krausfid}).

\subsection{Average over the Fubini-Study measure of the uncertainty of $L$} \label{apdx:haarweingarten}

First, rewriting $\int \text{d}\rho \Tr(\rho M^\dag M)$ yields
$$\int dU \, \Tr(U \rho U^\dagger M^\dagger M)
= \Tr\left[\left(
\int dU\, U\rho U^\dagger \right)M^\dagger M\right]$$,
where the integration is performed over the uniform Haar measure in the space of unitaries. Using the identity
$$
\int dU\, UX U^\dagger = \frac{\operatorname{Tr}(X) I}{d}$$
{valid} for any linear operator $X$, for the special case of $\rho$ pure one obtains
\begin{equation} \label{eq:haarmean} \int d \rho \,\, \text{Tr}(\rho M^\dagger M) = \frac{1}{d} \text{Tr}(M^\dagger M) . 
\end{equation}

Now, rewriting $\int \text{d}\rho \Tr(\rho M^\dag M)$ yields 
\begin{eqnarray}
&\int& dU \Tr(U\rho U^\dagger  M)^2 = \nonumber \\ 
\sum_{\substack{i,j,k,l \\ m,n,p,q}}&\int& dU\, U_{ij} U_{k\ell} \bar U_{mn} \bar U_{pq} \rho_{jn} \rho_{\ell q} M_{mi}  M_{pk} \nonumber .
\end{eqnarray}

\citet{collins_weingarten_2021} provide formulae to integrate polynomials of unitary matrices 
\begin{eqnarray} \label{eq:weingartendecomp}
\int_{U_d} dU U_{ij} U_{k\ell} \bar U_{mn}
\bar U_{pq} = \nonumber \\
\frac{1}{d^2-1}\left[(\delta_{im}\delta_{jn} \delta_{kp}\delta_{\ell q} + \delta_{ip}\delta_{jq} \delta_{km}\delta_{\ell n} ) {\color{white}\frac1d}\right. \nonumber  \\ 
{\color{white}\frac1d} - \left.\frac1d
(\delta_{im} \delta_{jq} \delta_{kp}\delta_{\ell n}
+\delta_{ip} \delta_{jn} \delta_{km}\delta_{\ell q})
\right] \;,
\end{eqnarray}
which contracting the indices gives
\begin{equation} 
\int d \rho \,\, \left|\text{Tr}(\rho M)\right|^2 = \frac{1}{d(d+1)} \left(\text{Tr}(M^\dagger M) + |\text{Tr}(M)|^2 \right). \nonumber
\end{equation} 

{Finally} subtracting (\ref{eq:weingartendecomp}) from (\ref{eq:haarmean}) {leads to} (\ref{eq:haarmeanfull}).

\section{Higher{-}order effects of the collapse operators}\label{apdx:exp}

\subsection{Complementary figure : deviation from linearity and gate dependence}\label{apdx:devlin}

\begin{figure}[!h]
    \centering
    \includegraphics[width=\columnwidth]{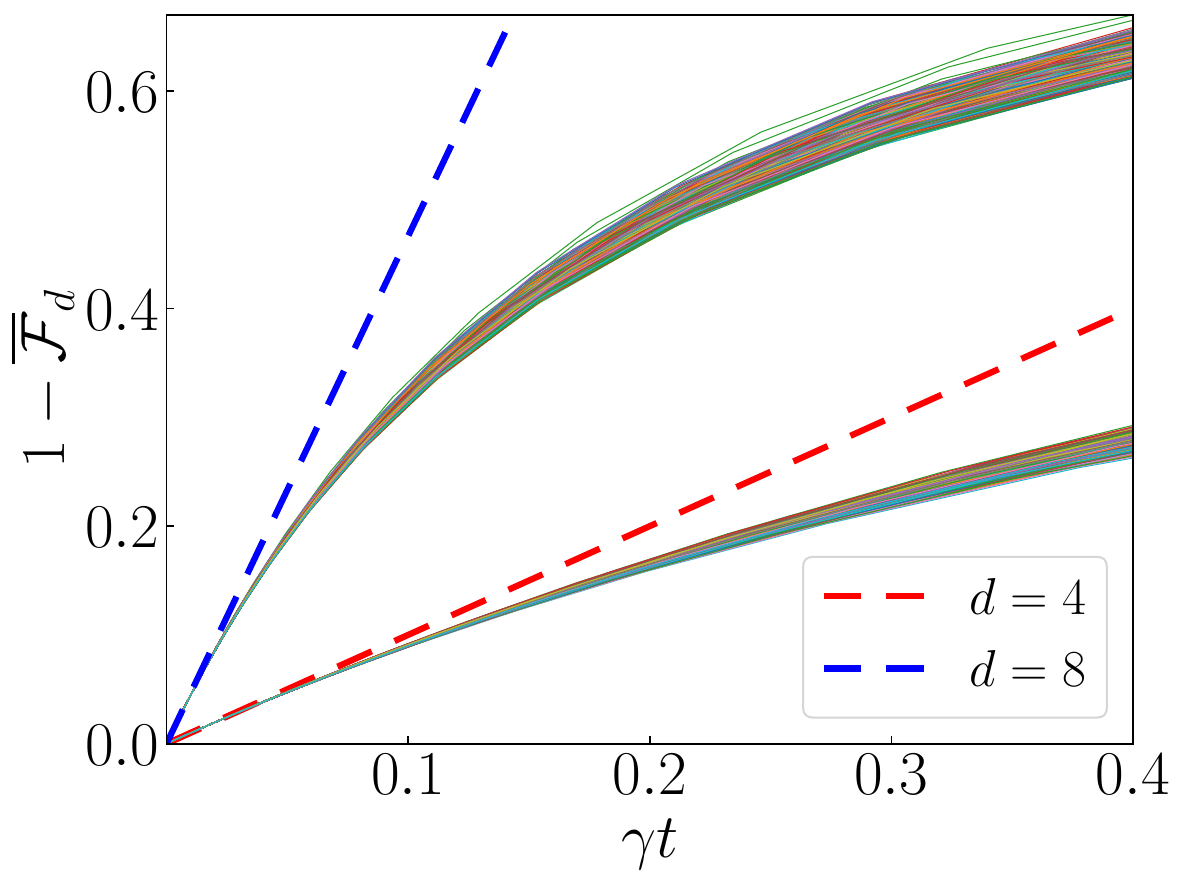}
    \caption{Simulated AGIs of $N_g=4400$ gates for $d=4,8$ in solid lines. The dashed lines correspond to the expected linear behaviour at small $\gamma t$.}
    \label{fig:gates48}
\end{figure}
Fig.\ref{fig:gates48} allows for observation of (i) the deviation from linear behaviour as $\gamma t$ increases, (ii) for higher $d$, this deviation becomes noticeable for smaller values of $\gamma t$ and, (iii) the infidelity becomes increasingly gate-dependent as $\gamma t$ increases.

\subsection{Full expansion {of the density matrix}}

{The density matrix} $\rho(t)$ can be decomposed as
\begin{equation}\label{eq:rholk}
    \rho(t) = \rho^{*} + {\sum_{l=1}\sum_{k=1} }\rho_{lk} \gamma^l t^k \;.
\end{equation}

Substituting Eq. (\ref{eq:rholk}) in (\ref{eq:master}) yields the following results:
\begin{itemize}
\item $\rho_{11} = \mathcal{D}[\rho^*] := \sum_k L_k\rho^* L_k^\dag -\frac12\{L_k^\dag L_k,\rho^*\}$\;.
\item for $l\geq2$, $k=1$, $\rho_{l,1} = 0$ \;.
\item for $l=1$, $k\geq2$,
\begin{equation}
    k \rho_{1k} = -i [H,\rho_{1(k-1)}] - \dot{\rho}_{1(k-1)} \;.
\end{equation}
\item $\forall l,k\geq2$, 
\begin{equation}
    k \rho_{lk} = -i [H,\rho_{l(k-1)}] - \dot\rho_{l(k-1)} + \mathcal{D}[\rho_{(l-1)(k-1)}]\;.
\end{equation}
\end{itemize}
It can be linked to (\ref{eq:rhopertresult}) by noticing that $M=-\rho_{11}$. 

Moreover, for $k=l=2$ {we obtain}
\begin{equation}\label{eq:(gammat)2}
     \rho_{22} = \mathcal{D}[\rho_{11}] = \mathcal{D}\left[\mathcal{D}[\rho^*]\right]\;.
\end{equation}

Finally {we have}
\begin{equation}
     \rho_{12} = \frac{i}{2}\left(\mathcal{D}\left[[H,\mathcal{\rho^*}]\right] - [H,\mathcal{\rho^*}]\right),
\end{equation}
{and}
\begin{equation}
     \rho_{13} = -\frac{i}{3}[H,\rho_{12}] - \frac{\dot\rho_{12}}{3}.
\end{equation}

This gives us the following expansion
\begin{equation}
    \label{eq:gammat2}\rho(t) = \rho^* + \gamma t \rho_{11} + \gamma t^2 \rho_{12} +\gamma t^3 \rho_{13} +(\gamma t)^2 \rho_{22} + \epsilon,
\end{equation}
with $\epsilon= O\left(\gamma^l t^k\right)_{l+k\geq5}$. 

Interestingly, it can be proven by induction that if $H=\mathbb{0}_d$ {then},
\begin{equation}
\label{eq:(gammat)k}\rho(t) = \rho^* + \sum_k (\gamma t)^k \rho_{kk},
\end{equation}
with $\rho_{kk} = \frac{1}{k!}\mathcal{D}^{(k)}[\rho^*]$.


\nocite{*}
\bibliography{bibliography}

\end{document}